\documentclass[aps,prl,twocolumn]{revtex4}
\usepackage{graphicx}
\usepackage{latexsym}
\usepackage{amsmath}
\usepackage{amsfonts}
\usepackage{amssymb}
\usepackage{color}

\usepackage{units}

\newcommand{\be}{\begin{equation}}
\newcommand{\ee}{\end{equation}}
\newcommand{\bea}{\begin{eqnarray}}
\newcommand{\eea}{\end{eqnarray}}
\newcommand{\bs}{\begin{split}}
\newcommand{\es}{\end{split}}
\newcommand{\bseq}{\begin{subequations}\begin{align}}

\begin{document}
\title{Activity induced phase separation}

\author{A.Y. Grosberg${}^{1,2}$ and J.-F. Joanny${}^{1,3}$ }
\affiliation{ ${}^1$ Physico-Chimie Curie UMR 168, Institut Curie, PSL Research
University, 26 rue d'Ulm, 75248 Paris Cedex 05, France\\ ${}^2$ Department of
Physics and Center for Soft Matter Research, New York University, 4 Washington
Place, New York, NY
10003, USA\\ ${}^3$ ESPCI-ParisTech, 10 rue Vauquelin 75005 Paris, France}
\date{\today}

\begin{abstract}
We consider a mixture of passive (i.e., Brownian) and active (e.g., bacterial or
colloidal swimmers) particles, and analyze the stability conditions of either
uniformly mixed or phase segregated steady states consisting of phases
enriched with different types of particles.  We show that in
sufficiently dilute mixtures the system behaves as if it were exposed to two
separate heat baths of uneven temperatures. It can be described within a second
virial approximation neglecting three body and higher order collisions.  In this
approximation, we define non-equilibrium ``chemical potentials'' whose gradients
govern diffusion fluxes and a non-equilibrium ``osmotic pressure'', which governs
the mechanical stability of the interface.
\end{abstract}

\maketitle

\textbf{Introduction.} Suspensions of actively moving particles, performing
mechanical work at the expense of internal or external energy consumption, have
attracted much attention over the last years \cite{CatesTailleur_EPL_2010,
Cates_group_PhysRevLett_2013, RednerHaganBaskaran_PhysRevLett_2013,
Palacci_Chaikin_Science_2013, Cates_group_NatureCommun_2014,
Cates_group_SoftMatter_2014, Hagan_group_SoftMatter_2014,
Marchetti_group_SoftMatter_2014, Palacci_Chaikin_2014}.  The interest is motivated
by biological applications, but these studies shed also
light on the fundamentals of statistical mechanics.  These systems share many
interesting properties such as spontaneous flows
\cite{RevModPhys_ActiveSoftMatter_2013}, but one of the most exciting phenomena is phase segregation \cite{CatesTailleur_EPL_2010,
Cates_group_PhysRevLett_2013, Hagan_group_SoftMatter_2014,
Marchetti_group_SoftMatter_2014, Palacci_Chaikin_2014}.  It is often driven by
variants of the so-called quorum sensing, which is the feedback mechanism reducing
the activity of a given active particle in the presence of a high concentration of
other active particles. Another
type of phase segregation can occur in a mixture of particles with different
levels of activity (active and non-active), when one phase is
enriched in active and the other one in passive particles
\cite{Cates_group_MixtureSegregation_ArXive_2014, Awazu_PhysRevE_2014}. This type
of active phase segregation is far less understood.

A new spin on the problem comes from Ref. \cite{GanaiSenguptaMenon_NAR_2014}.
These authors study eukaryotic nuclei and
the spatial segregation between eu- and hetero-chromatin, i.e., between actively
processed and almost silent parts of the genome.  Viewing the genome as a polymer,
they argue that genes which are being expressed
and, therefore, subject to RNA polymerization and other active processes, should
be viewed as active monomers, and silent genes are passive monomers.  They
then hypothesize that the observed compartmentalization between the two kinds of
chromatin is a phase segregation or rather a microphase segregation
\cite{Bates_FredricksonPhysicsToday_BlockCopolymers_1998}
based on activity. Their computational model appears reasonably consistent with
the data, even though the simulation replaces the ``activity''
by a sufficiently high effective temperature imposed on the active monomers by a
separate heat bath.  One striking observation is that the effective temperature
must be significantly
higher than the real temperature of the passive monomers (by about a factor
20).

In this work we develop a minimal analytical model of phase segregation between
active and passive particles, call them ${\cal{A}}$ and
${\cal{B}}$.  Our main idea is to look at systems of sufficiently low
concentration, where we resort to the type of reasoning
which for equilibriums systems leads to a virial expansion. We
give a systematic development of the second virial approximation in which only
pair collisions between particles are considered.

If the interactions are short range, at low concentration, each particle
completely looses its orientational correlations in the time between collisions
with other particles.  We can then
consider the particles
with two distinct levels of activity as exposed to two different heat baths, with
temperatures $T_{\cal{A}} \neq T_{\cal{B}}$.  This is clearly a system far from
equilibrium, with energy flowing from the hotter to the colder reservoir
via the interactions between particles; more physically, energy is
taken from the source of activity and dissipated into the surrounding medium via
our system of particles.  There are many examples of systems whose description
involves two distinct temperatures, ranging from plasmas (see basics in, e.g.,
\cite{LandauLifshitz_Kinetics}), to spin glasses \cite{Dotsenko_spin_glasses_book}
and heteropolymers \cite{RevModPhys.72.259}. Other examples are given in
\cite{Exartier199994, Khokhlov_TwoTemperatures_2004, PhysRevE.85.061127, DotsenkoMaciolekVasilyevOshanin_PRE_2013}.  In all these works (with the notable exception of Ref.
\cite{DotsenkoMaciolekVasilyevOshanin_PRE_2013}), the two temperatures are used to
describe motions on vastly different time scales.  In our system, there is no such
separation of time scales, and our
goal is to explore how the access to two different heat baths enhances the
tendency towards phase segregation.

Our starting point is the over-damped Langevin equation
\be \zeta_{i} \dot{x}_{i} = - \partial_{i} U + (2 T_{i} \zeta_{i})^{1/2} \xi_{i}
(t) \ , \label{eq:Langevin} \ee
for every particle $i$ in the system.  Here $x_i$ indicates the position of
particle
$i$ (for brevity, we will make no distinction between particles in 1 or 3
dimensions).  We assume that all forces acting on a particle derive from a
potential energy $U$, while $\partial_{i}$ is the derivative with respect
to $x_i$.  The friction coefficient for particle $i$ is $\zeta_{i}$, and $\xi_{i}
(t)$ is  a standard zero mean and unit variance Gaussian white noise, independent
for all particles $i$.  Finally, $T_{i}$ is the temperature of the heat bath interacting with particle
$i$, it is either $T_{\cal{A}}$ or $T_{\cal{B}}$.

A prototypical system that may be described by this model include mixtures of
actively swimming bacteria with either oxygen-starved ones (which do not actively
swim), or just similar sized inactive colloidal particles; another example is a
mixture of passive colloidal particles with the ones capable of light-induced
catalytic transformation of the solvent, like in
\cite{Palacci_Chaikin_Science_2013}.  In the jargon of the field
\cite{Many_authors_pressure_2014_1, Many_authors_pressure_2014_2}, these are
either run-and-tumble or active Brownian particles models, sufficiently dilute to
be amenable to a virial approximation.
Another major assumption behind our model is that we ignore hydrodynamic
interactions (Rouse model), each particle experiencing a local friction against an
immobile solvent.

\textbf{Two particles.} We first examine a system with only two particles, one ${\cal{A}}$ and
one ${\cal{B}}$.  In this case, Eq.(\ref{eq:Langevin}) consists of two coupled
Langevin equations. Dotsenko et al \cite{DotsenkoMaciolekVasilyevOshanin_PRE_2013}
studied this problem when the potential energy
$U(x_{\cal{A}},x_{\cal{B}})$ is a positive-definite quadratic form of its two
variables.  We only assume here that $U$ depends on
the distance between particles, $r=x_{\cal{A}}-x_{\cal{B}}$, i.e., $U =
u^{\cal{AB}}(r)$ and consider mostly cases where
$u^{\cal{AB}}(r)$ vanishes at large $r$.  We
derive a Langevin equation for the variable $r$, by combining
the two equations (\ref{eq:Langevin}):
\be \zeta_r \dot{r} = F(r) + (2 \zeta_r  \overline{T})^{1/2} \xi_r \ ,
\label{eq:Langevin_for_r} \ee
where the relevant friction is $\zeta_r=\zeta_{\cal A} \zeta_{\cal B} /
(\zeta_{\cal A} + \zeta_{\cal B})$ and the relevant temperature is the mobility-weighted average
$\overline{T} = (\zeta_{\cal B} T_{\cal A} +\zeta_{\cal A} T_{\cal B})/
(\zeta_{\cal
A} + \zeta_{\cal B})$.  The definition of the effective temperature $\overline{T}$
is dictated by the condition that the noise $\xi_r(t)$ is a zero mean, unit
variance Gaussian white
noise.  Since $\zeta_{{\cal A}, {\cal B}}>0$ are positive, $\overline{T}$ is always
between $T_{\cal{A}}$ and $T_{\cal{B}}$.  It follows from the
Langevin equation (\ref{eq:Langevin_for_r}) that the relative distance
between particles, $r$, in a steady state is Boltzmann distributed with the average temperature
$\overline{T}$, despite the fact that the system remains, of course, out of
equilibrium:
\be P(r) =  \exp \left( - U(r) / \overline{T} \right) z^{-1} \ , \label{eq:Boltzman_for_r} \ee
$z$ is here the ``partition sum'' ensuring normalization \footnote{See also Supplementary Material at URL ???  regarding enhancement of joint diffusion of two particles when $T_{{\cal A}} \neq T_{{\cal B}}$.}.

\textbf{Fokker-Planck equation, currents, and violation of detailed balance.}  The Langevin equations (\ref{eq:Langevin}) can be recast as a Fokker-Planck equation for the joint probability distribution of the coordinates of all particles $P(\{x \})$ and the corresponding currents $J_i$:
\begin{equation} \label{eq:FP}  \dot{P} = - \partial_i J_i, \quad
 J_i = - \partial_i U \  P / \zeta_i- T_i  \partial_i P / \zeta_i\ ,\end{equation}
(using the Einstein convention for the summation over repeated indices). At steady
state, for two particles the probability $P(r)$ depends only
on the distance $r = x_{\cal{A}}-x_{\cal{B}}$.  Then
$\partial_{\cal{A}} P = - \partial_{\cal{B}} P$, as well as $\partial_{\cal{A}} U
= - \partial_{\cal{B}} U$.  Furthermore, under these conditions, there is no
current in the $r = x_{\cal{A}} - x_{\cal{B}}$ direction. The current vector
$\mathbf{J}$ must be in the perpendicular direction, which means $J_{\cal{A}} -
J_{\cal{B}} = 0$, or
\be  \left( - \partial_{\cal{A}} U
/ \zeta_{\cal{A}} + \partial_{\cal{B}} U / \zeta_{\cal{B}} \right) - \left(
T_{\cal{A}} \partial_{\cal{A}} P / \zeta_{\cal{A}}- T_{\cal{B}}
 \partial_{\cal{B}} P / \zeta_{\cal{B}}\right) = 0 \ .
\nonumber \ee
This, of course, reproduces  the Boltzmann distribution with the average temperature
(\ref{eq:Boltzman_for_r}). We now compute the current $J_{\cal{A}}$, taking
advantage of 
the above result to replace the diffusion term:
\be J_{\cal{A}} = J_{\cal{B}} =  \frac{T_{\cal{A}} - T_{\cal{B}}}{T_{\cal{A}}
\zeta_{\cal{B}} + T_{\cal{B}} \zeta_{\cal{A}}}  P(r) \partial_{\cal{A}} U \ .
\label{eq:one_current} \ee
As expected, the current vanishes for an equilibrium system ($T_{\cal{A}} =
T_{\cal{B}}$), where detailed balance is obeyed.  However, if $T_{\cal{A}} \neq
T_{\cal{B}}$, detailed balance is violated.  For instance, if the
system is 1-dimensional, then its configuration space $(x_{\cal{A}},x_{\cal{B}})$
is a 2-dimensional plane, and it is easy to visualize loops of current
$\mathbf{J}$ in such a plane as shown on Fig. \ref{fig:Loopy_Currents}.  Physically, as
Eq (\ref{eq:one_current}) suggests, these loops of current mean that the more passive particle moves, on
average, mainly in the direction of the force acting on them, while the more
active particle moves in the direction opposite to the force acting on them.

\begin{figure}
  \centering
  \includegraphics[width=0.45\textwidth]{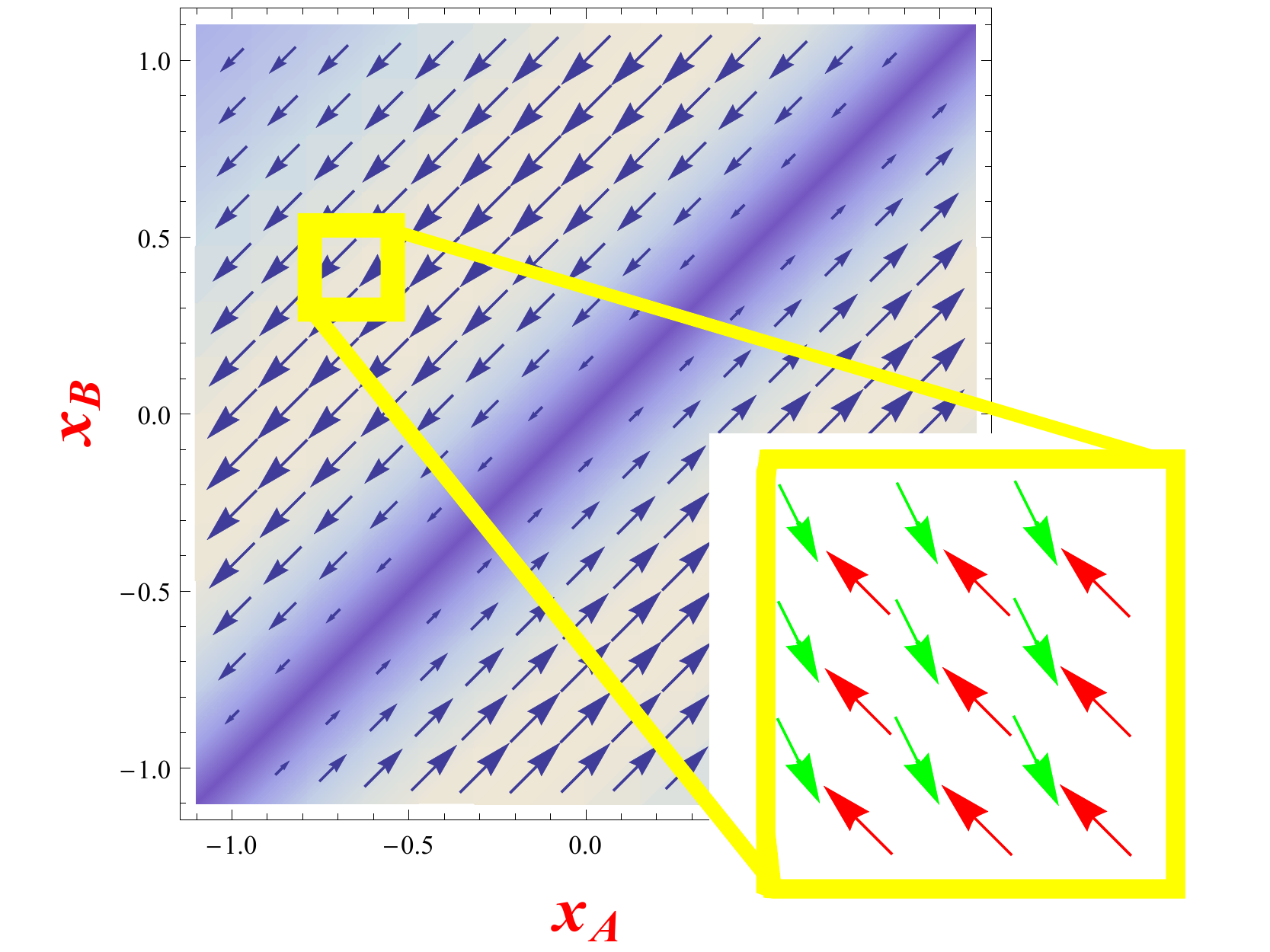}\\
  \caption{Field of currents.  For illustration purposes, the interaction potential is chosen in the form $u(r) =1/(r^2+1)$, and temperatures are $T_{\cal{A}} = 0.5$ and $T_{\cal{B}}=1$.  In the enlarged window, drift (upward and to the left) and diffusion (downward and to the right) are shown separately; unlike equilibrium system, these currents are not collinear, their vector sum gives rise to the non-potential current field. }\label{fig:Loopy_Currents}
\end{figure}

\textbf{Power transfer.}  As we pointed out above, energy is transferred
everywhere in the system from the ``hot heat bath'' to the ``cold'' bath, or from
the energy source of active motion to the surrounding passive medium.  The transferred power from ${\cal{B}}$ to ${\cal{A}}$ is the average $w = \left< - \dot{x}_{\cal{A}} \partial_{\cal{A}} U \right>$.  We note that velocity $\dot{x}_{\cal{A}}
= J_{\cal{A}}/P$, while the average involves an integration with a weight $P$.
Therefore, $w = - \int J_{\cal{A}} \partial_{\cal{A}} U d x_{\cal{A}}
dx_{\cal{B}}$, leading to:
\be w_{{\cal B}\to{\cal A}} = \frac{T_{\cal{B}} - T_{\cal{A}}}{T_{\cal{A}} \zeta_{\cal{B}} + T_{\cal{B}} \zeta_{\cal{A}}} \int \left( \partial_r u^{\cal{AB}}(r) \right)^2 \frac{e^{- u^{\cal{AB}}(r)/\overline{T}}}{z} d r \ . \label{eq:Dissipation} \ee
As expected, this power transfer from ${\cal B}$ to ${\cal A}$ vanishes for the equilibrium system if $T_{\cal{B}} = T_{\cal{A}}$, but it is positive if $T_{\cal{B}} > T_{\cal{A}}$ and negative otherwise.

To understand the meaning of the result (\ref{eq:Dissipation}), consider a 3-
dimensional system (when the  integration over $dr$ runs over the volume) and an
interaction potential $U$ that does not bind the particles together. An example is
a repulsive $U(r)$ with a little bump of energy scale $U_0$ and spatial scale
$\ell$.  Then the integral in Eq.(\ref{eq:Dissipation}) is estimated as
$(U_0/\ell)^2 \ell^3/L^3$, with $L^3$ the box volume (which enters in $z$), and
then (assuming for simplicity $\zeta_{\cal A}= \zeta_{\cal B} =\zeta$) the result
can be re-arranged as
\be w \sim \left[\ell \left( \overline{T}/\zeta \right)/{L^3} \right]
\left[\left(T_{\cal B} -T_{\cal A} \right) U_0^2 / \overline{T}^2 \right]\ . \ee
The first factor in the square brackets is the inverse Smoluchowski time between
collisions of two particles, and, therefore, the second factor is an estimate of
the energy transferred during one collision.

\textbf{Many particles.} Consider now a system of $N_{i}$
particles ${i}$ ($i,j = {{\cal A}}, {{\cal B}}$).  The Fokker-Planck equation (\ref{eq:FP}) is generalized
for any number of particles.  Integrating the Fokker-Planck equation over all
coordinates except for one, we derive a diffusion equation for the single particle
probability (proportional to the concentration) for every particle species:
\begin{equation} \begin{split}&  \frac{ \partial p_1^{{\cal A}}(\mathbf{r})}
{\partial t}  = \frac{N_{{\cal A}}}{\zeta_{{\cal A}}}
\partial_{\mathbf{r}} \left[ \int \frac{\partial u^{{\cal AA}}}{{\partial \mathbf{r}}}  p_2^{{\cal AA}} \left(
\mathbf{r}, \mathbf{r}^{\prime} \right) d \mathbf{r}^{\prime} \right] +  \\ & +
\frac{N_{{\cal B}}}{\zeta_{{\cal B}}} \partial_{\mathbf{r}} \left[ \int
\frac{\partial u^{{\cal AB}}}
{\partial {\mathbf{r}}} p_2^{{\cal AB}} \left( \mathbf{r}, \mathbf{r}^{\prime}
\right) d \mathbf{r}^{\prime} \right] +  \frac{T_{{\cal A}}}{\zeta_{{\cal A}}} \nabla^2_{\mathbf{r}}
p_1^{{\cal A}}(\mathbf{r})  \ . \label{eq:Equation_for _p1} \end{split}
\end{equation}
A similar equation is obtained for $p_1^{{\cal B}}(\mathbf{r})$.  These equations
for single particle probabilities include pair probabilities $p_2$.  By
integrating the multiparticle Fokker-Planck equation over all coordinates
except for two,
we derive equations for $p_2$, which include the 3-body correlations $p_3$,
and then a hierarchy of equations \footnote{See details about this hierarchy in Supplementary Material at URL ???}.  However, if the density is small enough, we
can neglect the 3-body correlations and ignore all terms involving $p_3$, thus
obtaining closed equation for $p_2$. Consistent with the two particle
system, we obtain in this approximation:
\begin{equation} \label{eq:ansatz_for_p2} p_2^{{ij}}(\mathbf{r},
\mathbf{r}^{\prime}) = p_1^{{i}}(\mathbf{r}) p_1^{{j}}(\mathbf{r}^{\prime}) \exp
\left[ - u^{{ij}} \left( \mathbf{r} - \mathbf{r}^{\prime} \right) / T_{{i}}
\right] \end{equation}
where $i,j ={\cal A}, {\cal B}$. The effective temperatures entering these expressions
is different for the three types of interactions: $T_{\cal AA}= T_{\cal A}$,
$T_{\cal BB}= T_{\cal B}$ and $T_{\cal AB}= \overline T$.

It is important to note that these distributions form only as a result of an
averaging over many collisions happening in the system under steady state
conditions (similar in this respect to an
equilibrium system).

Inserting the \textit{ansatz} (\ref{eq:ansatz_for_p2}) into equations
(\ref{eq:Equation_for _p1}) and introducing the concentrations $c^{{i}}
(\mathbf{r})=N_{{i}} p_1^{{i}}
(\mathbf{r})$, we obtain closed equations for the concentrations:
\begin{equation} \label{eq:Equations_for _p1_closed}  \frac{ \partial c^{{i}}
(\mathbf{r})}{\partial t} =  \frac{1}{\zeta_{{i}}} \frac{\partial}{\partial
\mathbf{r}} \left( c^{{i}} \frac{\partial \mu_{{i}} }{\partial
{\mathbf{r}}}\right) \end{equation}
These equations look like regular diffusion equations, but they are governed by
\textit{non-equilibrium} analogs of chemical potentials \footnote{See detailed derivation of non-equilibrium chemical potentials in Supplementary Material at URL ???}:
\begin{equation} \label{eq:non-eq_chemical_potentials}
\mu_{{\cal A}} = T_{{\cal A}} \ln c^{{\cal A}} + T_{{\cal A}} B_{{\cal A}}
c^{{\cal A}} +  \overline{T} B_{{\cal AB}} c^{{\cal B}}
\end{equation}
and a similar equation for $\mu_{{\cal B}}$. The virial coefficients are defined
each with its own temperature, as $B_{ij} = \int \left[1 - e^{ -u^{ij}
(\mathbf{r})/T_{ij} }\right]  d^3 \mathbf{r}$.

Non-equilibrium chemical potentials, as quantities whose gradient determines
the flux, were discussed in Ref. \cite{Cates_group_NatureCommun_2014}.  It was
shown, that, unlike its equilibrium counterpart, a non-equilibrium chemical
potential, in general, cannot be obtained as a derivative of a free energy.
This was shown in
particular for the gradient terms. In our
case, the situation is
different, because the non-equilibrium chemical potentials $\mu_{\cal{A}}$ and
$\mu_{\cal{B}}$ appear to be the partial derivatives $\mu_{{i}} = \frac{\partial
f}{\partial c^{{i}}}$ of a function, which
looks like a two-temperature free energy (per unit volume):
\be \begin{split} f & =  T_{{\cal A}} c^{{\cal A}} \ln \left( c^{{\cal A}} / e \right) +
T_{{\cal B}} c^{{\cal B}} \ln \left( c^{{\cal B}}/e \right) + \\ & +  (1/2) T_{{\cal
A}} B_{{\cal A}} c_{{\cal A}}^2 +  (1/2) T_{{\cal B}} B_{{\cal B}} c_{{\cal
B}}^2 +    \overline{T} B_{{\cal AB}} c^{{\cal A}} c^{{\cal B}} \ .
\label{eq:Two_T_free_energy} \end{split} \ee

\textbf{Instability of the uniform state and ``spinodal.''}  Suppose that
$c_0^{{\cal A}}$ and $c_0^{{\cal B}}$ are the averaged spatially uniform
concentrations of both components.  By introducing small space dependent
perturbations $c^{i}(\mathbf{r}) = c_0^{i} + \delta c^{i}
(\mathbf{r})$, we perform a linear stability analysis in the standard way.
This shows that an instability occurs macroscopically under the
condition
\be \frac{\phi_{{\cal A}}}{1+\phi_{{\cal A}}} \frac{\phi_{{\cal B}}}{1+\phi_{{\cal B}}} > \frac{T_{{\cal A}} T_{{\cal B}}}{\overline{T}} \frac{B_{{\cal AB}}^2} {B_{{\cal A}}B_{{\cal B}}} \ , \label{eq:spinodal_condition} \ee
where we have defined the volume fractions $c^{i} B_{i} = \phi_{i}$. In general, the virial coefficients depend on temperature in a complex way. A simple limit corresponds to
purely excluded volume interaction potentials such that the $B$'s do not depend on
temperature. We study this case in the following.

In the plane $\phi_{{\cal A}}$ and $\phi_{{\cal B}}$, the non-equilibrium
equivalent of the spinodal line (\ref{eq:spinodal_condition}) is a
hyperbola (see figure \ref{fig:Spinodal}).
The contrast between temperatures favors instability, it works in the same
direction as contrast between interactions.  But this instability, to have a
physical meaning, must occur at $\phi_{{\cal A}} <1$ and $\phi_{{\cal B}}<1$;
moreover, $\phi_{{\cal A}} + \phi_{{\cal B}} <1$.  For instance, consider the most
symmetric case of identical particles in all respects except driven by different
temperatures: $\zeta_{{\cal A}}=\zeta_{{\cal B}}$, $B_{{\cal A}}=B_{{\cal
B}}=B_{{\cal AB}}$.  In this case the spinodal line is in the
physical range $\phi_{{\cal A}}+\phi_{{\cal B}} <1$ if the ratio of the two
temperatures is outside the range $17-12\sqrt{2} \approx 0.029 < \frac{T_{{\cal
B}}}{T_{{\cal A}}} < 17+12\sqrt{2}\approx 34$ \footnote{See more general discussion of instability condition driven by contrasts of temperatures, frictions, and interactions in the Supplementary Material at URL ???}.
Thus, a numerically large temperature contrast is required to achieve an
instability by temperature difference alone.  This is somewhat consistent with
numerical observation of Ganai et al \cite{GanaiSenguptaMenon_NAR_2014}, as they
used $T_{{\cal B}}/T_{{\cal A}}=20$.

\begin{figure}
  \centering
  \includegraphics[width=0.3\textwidth]{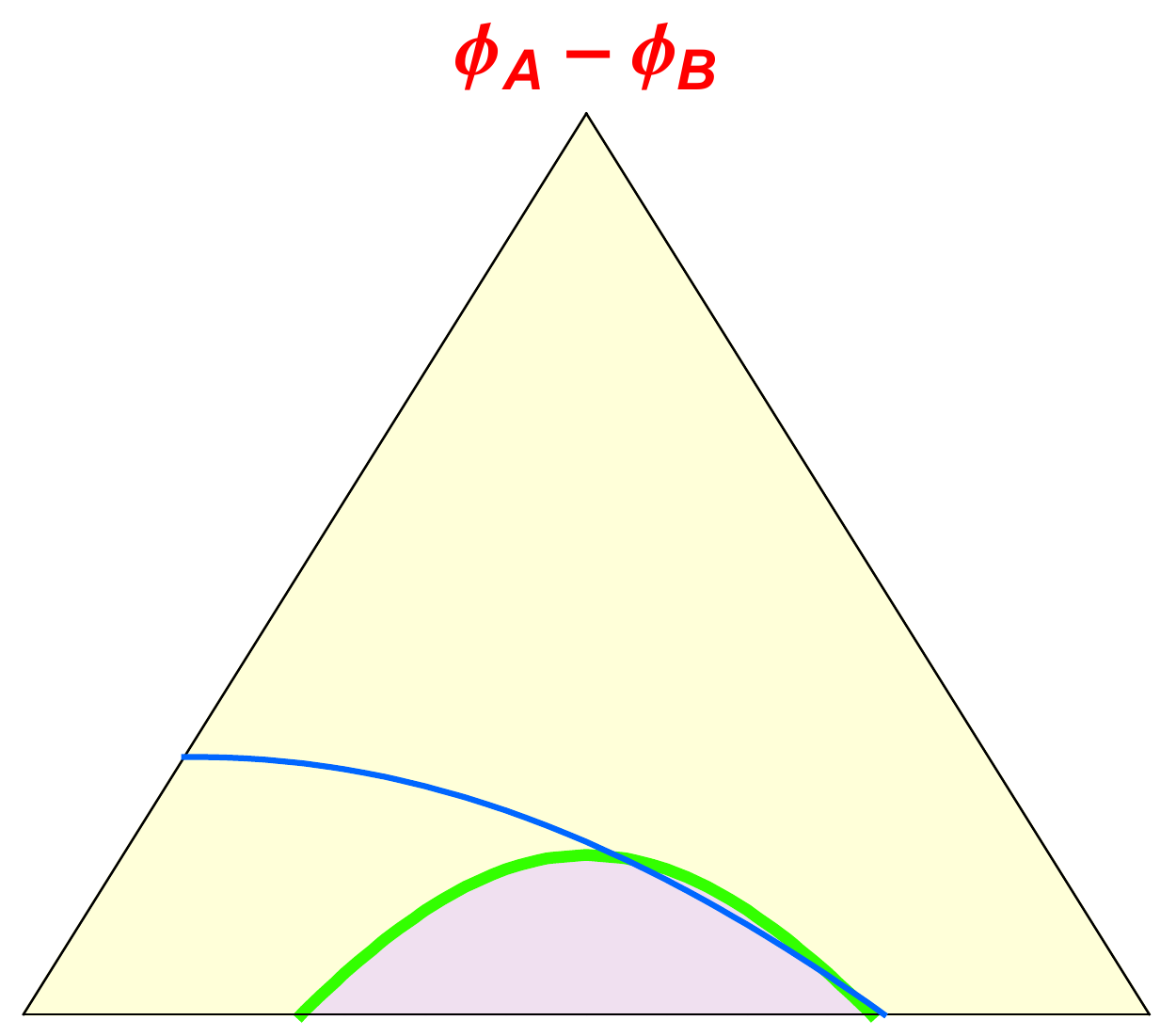}\\
  \caption{Phase diagram, $\phi_{\cal{A}}$ and $\phi_{\cal{B}}$ are volume fractions of ${\cal{A}}$- and ${\cal{B}}$-particles, while $1- \phi_{\cal{A}} - \phi_{\cal{B}}$ (which is the distance to lower side of the triangle) is the fraction of solvent.  Green line is the ``spinodal'' (below this line, uniformly mixed state is unstable).  Blue line corresponds to constant osmotic pressure, it shows the possibility of two states coexisting. }\label{fig:Spinodal}
\end{figure}

\textbf{Pressure and ``binodal.''} To address not only the loss of stability of
the uniform mixed state, but also the steady state phase segregation, in addition
to non-equilibrium chemical potentials we also need a non-equilibrium equivalent
of the osmotic pressure.  Given that non-equilibrium chemical potentials are the
derivatives of a ``quasi-free-energy'' (\ref{eq:Two_T_free_energy}), we can expect
the osmotic pressure to be given by Gibbs-Duhem formula $p = c^{{\cal A}} \mu_{{\cal
A}} + c^{{\cal B}} \mu_{{\cal B}} - f $, yielding
\be \begin{split} p & =  T_{{\cal A}} c^{{\cal A}} + T_{{\cal B}} c^{{\cal B}} +
\\ & +   (1/2) T_{{\cal A}} B_{{\cal A}} c_{{\cal A}}^2 + (1/2) T_{{\cal
B}} B_{{\cal B}} c_{{\cal B}}^2 +  \overline{T} B_{{\cal AB}} c^{{\cal A}}
c^{{\cal B}} \ . \label{eq:Two_T_pressure} \end{split} \ee
We have directly derived this result in two independent ways, first by computing
the force exerted on the wall by replacing the wall with a potential ramp, not
necessarily the same for both particle species, $u_1^{{\cal A}}(\mathbf{r})$ and
$u_1^{{\cal B}} \mathbf{r})$), and second by relating the pressure to the pair
correlation function in
the bulk \cite{Irving_Kirkwood_JCP_1950, BarratHansen_Liquids}.  Since we do not consider non-
spherical particles which experience a torque upon interactions with wall, we do
not have
the complications studied in the recent works
\cite{Takatori_Yan_Brady_PhysRevLett_2014, Many_authors_pressure_2014_1,
Many_authors_pressure_2014_2}, and, indeed, the two derivations \footnote{Details of both derivations are presented in Supplementary Material at URL ???} yield identical
results for the osmotic pressure (\ref{eq:Two_T_pressure}).  This result means
that
the densities of two coexisting phases at steady state are found by the Maxwell
common
tangent construction based upon the quasi-free energy function
(\ref{eq:Two_T_free_energy}).  So far, we make this statement in the second virial
approximation only.  It remains to be seen whether this is still true or not when
higher order collisions are taken into account, which is necessary at higher
densities.

The calculation of the dissipation (\ref{eq:Dissipation}), can be generalized to the dissipation per unit volume of the solution. The result is
\be w=c^{{\cal A}}c^{{\cal B}} \frac{T_{{\cal B}}-T_{{\cal A}}}{T_{{\cal
A}}\zeta_{{\cal B}} + T_{{\cal B}} \zeta_{{\cal A}}} \int \left( \frac{\partial
u^{{\cal AB}}}{\partial \mathbf{r}}\right)^2 e^{-
\frac{u^{{\cal AB}}(\mathbf{r})}{\overline{T}}} d^3 \mathbf{r} \ .
\label{dis}\ee
It shows that in a phase separated system the dissipation mostly happens around the phase boundary \footnote{Interestingly, more detailed derivation of the dissipated power shows that, in the system with pairwise interactions only, the dissipated power is exactly expressed using two- and three-body correlations only, as shown in Supplementary Material at URL ???}.

\textbf{Conclusion and discussion.}  To conclude, we first have to estimate the
temperatures $T_{{\cal A}}$ and $T_{{\cal B}}$ in terms of real parameters of
active particles.  Since we assumed that the re-orientation time of one particle
is much smaller than the time between collisions: $\tau_r \ll \tau_c$, the particle
trajectory between collisions is that of a random walk, characterized by an effective
diffusion constant $D_{\mathrm{eff}} = T_{\mathrm{eff}}/\zeta \simeq  v_0^2 \tau_r/6
+ D$, where $v_0$ is the swimming speed, $D=T/\zeta$ is the passive diffusion
constant, $T = T_{{\cal B}}$ is the real ambient temperature, and $T_{\mathrm{eff}}=T_{\cal A}$
corresponds in our theory to the temperature of the hotter heat bath.
The temperature difference, which controls dissipation rate
(\ref{dis}), is therefore given by $T_{{\cal A}} -
T_{{\cal B}} \simeq v_0^2\tau_r \zeta /6$, it is directly related to the level of activity measured by the swimming speed $v_0$.

Returning to the applicability condition $\tau_r \ll \tau_c$, for particles of
size $b$ with typical distance between particles $d$, the collision time is
estimated by the Smoluchowski formula $\tau_c \sim D_{\mathrm{eff}}b/d^3 \sim
D_{\mathrm{eff}} \phi/b^2$, where $\phi \sim b^3/d^3$ is the volume fraction of
particles.  The condition of applicability is then conveniently
formulated in terms of the Peclet number $v_0 \tau_r / b \equiv \mathrm{Pe} \ll 1/\sqrt{\phi}$. Thus, our theory should work if the system is dilute enough and/or the active drive is not too strong.

Another significant limitation of our approach is the fact that we neglect hydrodynamic
interactions.  This might be particularly important in the case of actively
swimming colloids, as they usually drive themselves by creating and maintaining a
train of diffusing chemicals, and accordingly their interaction upon approach and
collision is hardly describable in terms of a conservative force potential, as we
did here.

Despite all limitations, we believe that the theory developed here is useful
because it is physically transparent and may be instructive as a source of
physical intuition for these highly unusual driven systems.

\textbf{Acknowledgements}  This work was performed in Paris, where AYG was on a
long term visit.  AYG acknowledges the hospitality of both the Curie Institute
and ESPCI.
We thank M.E. Cates for a stimulating discussion.

\bibliography{Two_Temperatures_References}

\begin{thebibliography}{26}
\expandafter\ifx\csname natexlab\endcsname\relax\def\natexlab#1{#1}\fi
\expandafter\ifx\csname bibnamefont\endcsname\relax
  \def\bibnamefont#1{#1}\fi
\expandafter\ifx\csname bibfnamefont\endcsname\relax
  \def\bibfnamefont#1{#1}\fi
\expandafter\ifx\csname citenamefont\endcsname\relax
  \def\citenamefont#1{#1}\fi
\expandafter\ifx\csname url\endcsname\relax
  \def\url#1{\texttt{#1}}\fi
\expandafter\ifx\csname urlprefix\endcsname\relax\def\urlprefix{URL }\fi
\providecommand{\bibinfo}[2]{#2}
\providecommand{\eprint}[2][]{\url{#2}}

\bibitem[{\citenamefont{Cates and Tailleur}(2013)}]{CatesTailleur_EPL_2010}
\bibinfo{author}{\bibfnamefont{M.~E.} \bibnamefont{Cates}} \bibnamefont{and}
  \bibinfo{author}{\bibfnamefont{J.}~\bibnamefont{Tailleur}},
  \bibinfo{journal}{EPL (Europhysics Letters)} \textbf{\bibinfo{volume}{101}},
  \bibinfo{pages}{20010} (\bibinfo{year}{2013}).

\bibitem[{\citenamefont{Stenhammar et~al.}(2013)\citenamefont{Stenhammar,
  Tiribocchi, Allen, Marenduzzo, and E.Cates}}]{Cates_group_PhysRevLett_2013}
\bibinfo{author}{\bibfnamefont{J.}~\bibnamefont{Stenhammar}},
  \bibinfo{author}{\bibfnamefont{A.}~\bibnamefont{Tiribocchi}},
  \bibinfo{author}{\bibfnamefont{R.}~\bibnamefont{Allen}},
  \bibinfo{author}{\bibfnamefont{D.}~\bibnamefont{Marenduzzo}},
  \bibnamefont{and} \bibinfo{author}{\bibfnamefont{M.}~\bibnamefont{E.Cates}},
  \bibinfo{journal}{Phys. Rev. Lett.} \textbf{\bibinfo{volume}{111}},
  \bibinfo{pages}{145702} (\bibinfo{year}{2013}).

\bibitem[{\citenamefont{Redner et~al.}(2013)\citenamefont{Redner, Hagan, and
  Baskaran}}]{RednerHaganBaskaran_PhysRevLett_2013}
\bibinfo{author}{\bibfnamefont{G.}~\bibnamefont{Redner}},
  \bibinfo{author}{\bibfnamefont{M.}~\bibnamefont{Hagan}}, \bibnamefont{and}
  \bibinfo{author}{\bibfnamefont{A.}~\bibnamefont{Baskaran}},
  \bibinfo{journal}{Phys. Rev. Lett.} \textbf{\bibinfo{volume}{110}},
  \bibinfo{pages}{055701} (\bibinfo{year}{2013}).

\bibitem[{\citenamefont{Palacci et~al.}(2013)\citenamefont{Palacci, Sacanna,
  Steinberg, Pine, and Chaikin}}]{Palacci_Chaikin_Science_2013}
\bibinfo{author}{\bibfnamefont{J.}~\bibnamefont{Palacci}},
  \bibinfo{author}{\bibfnamefont{S.}~\bibnamefont{Sacanna}},
  \bibinfo{author}{\bibfnamefont{A.~P.} \bibnamefont{Steinberg}},
  \bibinfo{author}{\bibfnamefont{D.~J.} \bibnamefont{Pine}}, \bibnamefont{and}
  \bibinfo{author}{\bibfnamefont{P.~M.} \bibnamefont{Chaikin}},
  \bibinfo{journal}{Science} \textbf{\bibinfo{volume}{339}},
  \bibinfo{pages}{936} (\bibinfo{year}{2013}).

\bibitem[{\citenamefont{Wittkowski et~al.}(2014)\citenamefont{Wittkowski,
  Tiribocchi, Stenhammar, Allen, Marenduzzo, and
  Cates}}]{Cates_group_NatureCommun_2014}
\bibinfo{author}{\bibfnamefont{R.}~\bibnamefont{Wittkowski}},
  \bibinfo{author}{\bibfnamefont{A.}~\bibnamefont{Tiribocchi}},
  \bibinfo{author}{\bibfnamefont{J.}~\bibnamefont{Stenhammar}},
  \bibinfo{author}{\bibfnamefont{R.~J.} \bibnamefont{Allen}},
  \bibinfo{author}{\bibfnamefont{D.}~\bibnamefont{Marenduzzo}},
  \bibnamefont{and} \bibinfo{author}{\bibfnamefont{M.~E.} \bibnamefont{Cates}},
  \bibinfo{journal}{Nature Communications} \textbf{\bibinfo{volume}{5}},
  \bibinfo{pages}{4351} (\bibinfo{year}{2014}).

\bibitem[{\citenamefont{Stenhammar
  et~al.}(2014{\natexlab{a}})\citenamefont{Stenhammar, Marenduzzo, Allen, and
  Cates}}]{Cates_group_SoftMatter_2014}
\bibinfo{author}{\bibfnamefont{J.}~\bibnamefont{Stenhammar}},
  \bibinfo{author}{\bibfnamefont{D.}~\bibnamefont{Marenduzzo}},
  \bibinfo{author}{\bibfnamefont{R.~J.} \bibnamefont{Allen}}, \bibnamefont{and}
  \bibinfo{author}{\bibfnamefont{M.~E.} \bibnamefont{Cates}},
  \bibinfo{journal}{Soft Matter} \textbf{\bibinfo{volume}{10}},
  \bibinfo{pages}{1489} (\bibinfo{year}{2014}{\natexlab{a}}).

\bibitem[{\citenamefont{Fily et~al.}(2014)\citenamefont{Fily, Baskaran, and
  Hagan}}]{Hagan_group_SoftMatter_2014}
\bibinfo{author}{\bibfnamefont{Y.}~\bibnamefont{Fily}},
  \bibinfo{author}{\bibfnamefont{A.}~\bibnamefont{Baskaran}}, \bibnamefont{and}
  \bibinfo{author}{\bibfnamefont{M.~F.} \bibnamefont{Hagan}},
  \bibinfo{journal}{Soft Matter} \textbf{\bibinfo{volume}{10}},
  \bibinfo{pages}{5609} (\bibinfo{year}{2014}).

\bibitem[{\citenamefont{Yang et~al.}(2014)\citenamefont{Yang, Manning, and
  Marchetti}}]{Marchetti_group_SoftMatter_2014}
\bibinfo{author}{\bibfnamefont{X.}~\bibnamefont{Yang}},
  \bibinfo{author}{\bibfnamefont{M.~L.} \bibnamefont{Manning}},
  \bibnamefont{and} \bibinfo{author}{\bibfnamefont{M.~C.}
  \bibnamefont{Marchetti}}, \bibinfo{journal}{Soft Matter}
  \textbf{\bibinfo{volume}{10}}, \bibinfo{pages}{6477} (\bibinfo{year}{2014}).

\bibitem[{\citenamefont{Palacci et~al.}(2014)\citenamefont{Palacci, Sacanna,
  S.H.~Kim, and Chaikin}}]{Palacci_Chaikin_2014}
\bibinfo{author}{\bibfnamefont{J.}~\bibnamefont{Palacci}},
  \bibinfo{author}{\bibfnamefont{S.}~\bibnamefont{Sacanna}},
  \bibinfo{author}{\bibfnamefont{D.~J.~P.} \bibnamefont{S.H.~Kim},
  \bibfnamefont{G.R.~Yi}}, \bibnamefont{and}
  \bibinfo{author}{\bibfnamefont{P.~M.} \bibnamefont{Chaikin}},
  \bibinfo{journal}{Phil. Trans. Royal Soc. A: Math., Phys.}
  \textbf{\bibinfo{volume}{372}}, \bibinfo{pages}{20130372}
  (\bibinfo{year}{2014}).

\bibitem[{\citenamefont{Marchetti et~al.}(2013)\citenamefont{Marchetti, Joanny,
  Ramaswamy, Liverpool, Prost, Rao, and
  Simha}}]{RevModPhys_ActiveSoftMatter_2013}
\bibinfo{author}{\bibfnamefont{M.}~\bibnamefont{Marchetti}},
  \bibinfo{author}{\bibfnamefont{J.-F.} \bibnamefont{Joanny}},
  \bibinfo{author}{\bibfnamefont{S.}~\bibnamefont{Ramaswamy}},
  \bibinfo{author}{\bibfnamefont{T.}~\bibnamefont{Liverpool}},
  \bibinfo{author}{\bibfnamefont{J.}~\bibnamefont{Prost}},
  \bibinfo{author}{\bibfnamefont{M.}~\bibnamefont{Rao}}, \bibnamefont{and}
  \bibinfo{author}{\bibfnamefont{R.~A.} \bibnamefont{Simha}},
  \bibinfo{journal}{Rev. Mod. Phys.} \textbf{\bibinfo{volume}{85}},
  \bibinfo{pages}{1143} (\bibinfo{year}{2013}).

\bibitem[{\citenamefont{Stenhammar
  et~al.}(2014{\natexlab{b}})\citenamefont{Stenhammar, Wittkowski, Marenduzzo,
  and Cates}}]{Cates_group_MixtureSegregation_ArXive_2014}
\bibinfo{author}{\bibfnamefont{J.}~\bibnamefont{Stenhammar}},
  \bibinfo{author}{\bibfnamefont{R.}~\bibnamefont{Wittkowski}},
  \bibinfo{author}{\bibfnamefont{D.}~\bibnamefont{Marenduzzo}},
  \bibnamefont{and} \bibinfo{author}{\bibfnamefont{M.~E.} \bibnamefont{Cates}},
  \bibinfo{journal}{arXiv:1408.5175v1}  (\bibinfo{year}{2014}{\natexlab{b}}).

\bibitem[{\citenamefont{Awazu}(2014)}]{Awazu_PhysRevE_2014}
\bibinfo{author}{\bibfnamefont{A.}~\bibnamefont{Awazu}},
  \bibinfo{journal}{Phys. Rev. E} \textbf{\bibinfo{volume}{90}},
  \bibinfo{pages}{042308} (\bibinfo{year}{2014}).

\bibitem[{\citenamefont{Ganai et~al.}(2014)\citenamefont{Ganai, Sengupta, and
  Menon}}]{GanaiSenguptaMenon_NAR_2014}
\bibinfo{author}{\bibfnamefont{N.}~\bibnamefont{Ganai}},
  \bibinfo{author}{\bibfnamefont{S.}~\bibnamefont{Sengupta}}, \bibnamefont{and}
  \bibinfo{author}{\bibfnamefont{G.~I.} \bibnamefont{Menon}},
  \bibinfo{journal}{Nucleic Acids Research} \textbf{\bibinfo{volume}{42}},
  \bibinfo{pages}{4145} (\bibinfo{year}{2014}).

\bibitem[{\citenamefont{Bates and
  Fredrickson}(2008)}]{Bates_FredricksonPhysicsToday_BlockCopolymers_1998}
\bibinfo{author}{\bibfnamefont{F.~S.} \bibnamefont{Bates}} \bibnamefont{and}
  \bibinfo{author}{\bibfnamefont{G.~H.} \bibnamefont{Fredrickson}},
  \bibinfo{journal}{Physics Today} \textbf{\bibinfo{volume}{52}},
  \bibinfo{pages}{32} (\bibinfo{year}{2008}).

\bibitem[{\citenamefont{Lifshitz and
  Pitaevskii}(2002)}]{LandauLifshitz_Kinetics}
\bibinfo{author}{\bibfnamefont{E.~M.} \bibnamefont{Lifshitz}} \bibnamefont{and}
  \bibinfo{author}{\bibfnamefont{L.~P.} \bibnamefont{Pitaevskii}},
  \emph{\bibinfo{title}{Physical Kinetics (Course of Theoretical Physics,
  Volume 10)}} (\bibinfo{publisher}{Butterworth-Heinemann},
  \bibinfo{year}{2002}).

\bibitem[{\citenamefont{Dotsenko}(1994)}]{Dotsenko_spin_glasses_book}
\bibinfo{author}{\bibfnamefont{V.~S.} \bibnamefont{Dotsenko}},
  \emph{\bibinfo{title}{Introduction to the Theory of Spin Glasses and Neural
  Networks}} (\bibinfo{publisher}{World Scientific, Singapore},
  \bibinfo{year}{1994}).

\bibitem[{\citenamefont{Pande et~al.}(2000)\citenamefont{Pande, Grosberg, and
  Tanaka}}]{RevModPhys.72.259}
\bibinfo{author}{\bibfnamefont{V.~S.} \bibnamefont{Pande}},
  \bibinfo{author}{\bibfnamefont{A.~Y.} \bibnamefont{Grosberg}},
  \bibnamefont{and} \bibinfo{author}{\bibfnamefont{T.}~\bibnamefont{Tanaka}},
  \bibinfo{journal}{Rev. Mod. Phys.} \textbf{\bibinfo{volume}{72}},
  \bibinfo{pages}{259} (\bibinfo{year}{2000}).

\bibitem[{\citenamefont{Exartier and Peliti}(1999)}]{Exartier199994}
\bibinfo{author}{\bibfnamefont{R.}~\bibnamefont{Exartier}} \bibnamefont{and}
  \bibinfo{author}{\bibfnamefont{L.}~\bibnamefont{Peliti}},
  \bibinfo{journal}{Physics Letters A} \textbf{\bibinfo{volume}{261}},
  \bibinfo{pages}{94 } (\bibinfo{year}{1999}).

\bibitem[{\citenamefont{Chertovich et~al.}(2004)\citenamefont{Chertovich,
  Govorun, Ivanov, Khalatur, and Khokhlov}}]{Khokhlov_TwoTemperatures_2004}
\bibinfo{author}{\bibfnamefont{A.~V.} \bibnamefont{Chertovich}},
  \bibinfo{author}{\bibfnamefont{E.~N.} \bibnamefont{Govorun}},
  \bibinfo{author}{\bibfnamefont{V.~A.} \bibnamefont{Ivanov}},
  \bibinfo{author}{\bibfnamefont{P.~G.} \bibnamefont{Khalatur}},
  \bibnamefont{and} \bibinfo{author}{\bibfnamefont{A.~R.}
  \bibnamefont{Khokhlov}}, \bibinfo{journal}{The European Physical Journ. E}
  \textbf{\bibinfo{volume}{13}}, \bibinfo{pages}{15–25}
  (\bibinfo{year}{2004}).

\bibitem[{\citenamefont{Crisanti et~al.}(2012)\citenamefont{Crisanti, Puglisi,
  and Villamaina}}]{PhysRevE.85.061127}
\bibinfo{author}{\bibfnamefont{A.}~\bibnamefont{Crisanti}},
  \bibinfo{author}{\bibfnamefont{A.}~\bibnamefont{Puglisi}}, \bibnamefont{and}
  \bibinfo{author}{\bibfnamefont{D.}~\bibnamefont{Villamaina}},
  \bibinfo{journal}{Phys. Rev. E} \textbf{\bibinfo{volume}{85}},
  \bibinfo{pages}{061127} (\bibinfo{year}{2012}).

\bibitem[{\citenamefont{Dotsenko et~al.}(2013)\citenamefont{Dotsenko,
  Macio\l{}ek, Vasilyev, and
  Oshanin}}]{DotsenkoMaciolekVasilyevOshanin_PRE_2013}
\bibinfo{author}{\bibfnamefont{V.}~\bibnamefont{Dotsenko}},
  \bibinfo{author}{\bibfnamefont{A.}~\bibnamefont{Macio\l{}ek}},
  \bibinfo{author}{\bibfnamefont{O.}~\bibnamefont{Vasilyev}}, \bibnamefont{and}
  \bibinfo{author}{\bibfnamefont{G.}~\bibnamefont{Oshanin}},
  \bibinfo{journal}{Phys. Rev. E} \textbf{\bibinfo{volume}{87}},
  \bibinfo{pages}{062130} (\bibinfo{year}{2013}).

\bibitem[{\citenamefont{Solon et~al.}(2014{\natexlab{a}})\citenamefont{Solon,
  Fily, Baskaran, Cates, Kafri, Kardar, and
  Tailleur}}]{Many_authors_pressure_2014_1}
\bibinfo{author}{\bibfnamefont{A.~P.} \bibnamefont{Solon}},
  \bibinfo{author}{\bibfnamefont{Y.}~\bibnamefont{Fily}},
  \bibinfo{author}{\bibfnamefont{A.}~\bibnamefont{Baskaran}},
  \bibinfo{author}{\bibfnamefont{M.~E.} \bibnamefont{Cates}},
  \bibinfo{author}{\bibfnamefont{Y.}~\bibnamefont{Kafri}},
  \bibinfo{author}{\bibfnamefont{M.}~\bibnamefont{Kardar}}, \bibnamefont{and}
  \bibinfo{author}{\bibfnamefont{J.}~\bibnamefont{Tailleur}},
  \bibinfo{journal}{arXiv:1412.3952}  (\bibinfo{year}{2014}{\natexlab{a}}).

\bibitem[{\citenamefont{Solon et~al.}(2014{\natexlab{b}})\citenamefont{Solon,
  Stenhammar, Wittkowski, Kardar, Kafri, Cates, and
  Tailleur}}]{Many_authors_pressure_2014_2}
\bibinfo{author}{\bibfnamefont{A.~P.} \bibnamefont{Solon}},
  \bibinfo{author}{\bibfnamefont{J.}~\bibnamefont{Stenhammar}},
  \bibinfo{author}{\bibfnamefont{R.}~\bibnamefont{Wittkowski}},
  \bibinfo{author}{\bibfnamefont{M.}~\bibnamefont{Kardar}},
  \bibinfo{author}{\bibfnamefont{Y.}~\bibnamefont{Kafri}},
  \bibinfo{author}{\bibfnamefont{M.~E.} \bibnamefont{Cates}}, \bibnamefont{and}
  \bibinfo{author}{\bibfnamefont{J.}~\bibnamefont{Tailleur}},
  \bibinfo{journal}{arXiv:1412.5475}  (\bibinfo{year}{2014}{\natexlab{b}}).

\bibitem[{\citenamefont{Irving and Kirkwood}(1950)}]{Irving_Kirkwood_JCP_1950}
\bibinfo{author}{\bibfnamefont{J.~H.} \bibnamefont{Irving}} \bibnamefont{and}
  \bibinfo{author}{\bibfnamefont{J.~G.} \bibnamefont{Kirkwood}},
  \bibinfo{journal}{The Journal of Chemical Physics}
  \textbf{\bibinfo{volume}{18}}, \bibinfo{pages}{817} (\bibinfo{year}{1950}).

\bibitem[{\citenamefont{Barrat and Hansen}(2003)}]{BarratHansen_Liquids}
\bibinfo{author}{\bibfnamefont{J.-L.} \bibnamefont{Barrat}} \bibnamefont{and}
  \bibinfo{author}{\bibfnamefont{J.-P.} \bibnamefont{Hansen}},
  \emph{\bibinfo{title}{Basic concepts for simple and complex liquids}}
  (\bibinfo{publisher}{Cambridge University Press}, \bibinfo{year}{2003}).

\bibitem[{\citenamefont{Takatori et~al.}(2014)\citenamefont{Takatori, Yan, and
  Brady}}]{Takatori_Yan_Brady_PhysRevLett_2014}
\bibinfo{author}{\bibfnamefont{S.~C.} \bibnamefont{Takatori}},
  \bibinfo{author}{\bibfnamefont{W.}~\bibnamefont{Yan}}, \bibnamefont{and}
  \bibinfo{author}{\bibfnamefont{J.~F.} \bibnamefont{Brady}},
  \bibinfo{journal}{Phys. Rev. Lett.} \textbf{\bibinfo{volume}{113}},
  \bibinfo{pages}{028103} (\bibinfo{year}{2014}).

\end{thebibliography}

\newpage

\appendix

\begin{widetext}

\section{Activity induced phase separation:  Supplementary material}

In this Supplementary Material, we provide the following: \begin{enumerate} \item For two particles, we analyze their joint diffusion and show how it is enhanced by the uneven driving temperatures.  \item For many particles, we provide a detailed derivation of the hierarchy of the equations for the correlation functions and show how our second virial approximation comes out.  \item We provide detailed analysis of the instability conditions when not only temperatures are uneven, but so also frictions and interactions.  \item We provide two methods to derive the osmotic pressure.  \item We derive a more general expression for the transfer of power between the heat reservoirs that includes two-and three-body correlations. \end{enumerate} \end{widetext}

\maketitle

\section{Two particles: ``Center of friction'' diffusion .}

In the main text, we analyzed the relative motion of two particles ${\cal A}$ and
${\cal B}$ by looking at the variable $r = x_{{\cal A}} - x_{{\cal B}}$.  It is
interesting to find also how the presence of two distinct temperatures affects
their joint diffusion in space.  To do so, it is convenient to define their joint
coordinate $R$ in such a way that the noises in $r$ and $R$ are statistically
independent.  This is achieved by choosing
\be R = \frac{\zeta_{{\cal A}} T_{{\cal B}}}{\zeta_{{\cal A}} T_{{\cal B}} +
\zeta_{{\cal B}} T_{{\cal A}}} x_{{\cal A}} + \frac{\zeta_{{\cal B}} T_{{\cal A}}}
{\zeta_{{\cal A}} T_{{\cal B}} + \zeta_{{\cal B}} T_{{\cal A}}} x_{{\cal B}} \ .
\ee
Then, the Langevin equation for $R$ reads
\be \begin{split} \left( \zeta_{{\cal A}} + \zeta_{{\cal B}} \right) \dot{R} = &
\frac{ T_{{\cal B}} - T_{{\cal A}}}{\overline{T}} F(r) + \\ & + \sqrt{2 \left(
\zeta_{{\cal A}} + \zeta_{{\cal B}} \right) \frac{( T_{{\cal A}}  T_{{\cal B}}}
{\overline{T}}} \xi_R(t) \ , \end{split} \label{eq:Langevin_for_R} \ee
where $\xi_R(t)$ is a zero mean unit variance Gaussian white noise (independent of
$\xi_r$, as stated), and $\overline{T} = \left( \zeta_{{\cal A}} T_{{\cal B}} +
\zeta_{{\cal B}} T_{{\cal A}} \right)/\left( \zeta_{{\cal A}} + \zeta_{{\cal B}}
\right)$ was defined in the main text.  In an equilibrium system, at $T_{{\cal A}}
= T_{{\cal B}} $, the inter-particle force $F(r)$ does not couple to the joint
motion.   Not so out of equilibrium:  since on average the force $F(r)$ vanishes
(by
symmetry), $\left< F(r) \right>=0$, the $F(r)$ term provides an additional noise
driving the diffusion of the variable $R$.  Of course, it is not a white noise, so
that the
dynamics of $R$ is not a simple diffusion on time scales shorter or comparable
to the correlation time of $r$.  But on longer time scales $R$ undergoes simple
diffusion, with a diffusion coefficient which can be directly read out of the
Langevin equation (\ref{eq:Langevin_for_R}), because $F(r(t))$ is statistically
independent from $\xi_R(t)$:
\be  D_R  =  \frac{T_{{\cal A}}T_{{\cal B}}}{\zeta_{{\cal A}} T_{{\cal B}}
+\zeta_{{\cal B}} T_{{\cal A}}} +  \frac{1}{2} \left( \frac{T_{{\cal A}} -
T_{{\cal B}}}{\zeta_{{\cal A}} T_{{\cal B}} + \zeta_{{\cal B}} T_{{\cal A}}}
\right)^2 \left(F^2 \right)_{\omega = 0} \ ;  \ee
here the power the spectrum of  the force $F(r)$ at zero frequency is
\be \left(F^2 \right)_{\omega = 0} = \int_{-\infty}^{\infty} \left< F(r(t))
F(r(t+\tau)) \right> d \tau \ . \ee

in order to find the power spectrum of the force, we Fourier transform the
Langevin equation for $r$ (Eq.(2) of the main text):
\be F_{\omega} = \imath \omega \zeta_r r_{\omega} - \left( 2 \zeta_r \overline{T}
\right)^{1/2} \xi_{\omega} \ , \ee
Then multiplying it by the complex conjugate and assuming $\left[\omega F_{\omega}
\right]_{\omega = 0} = 0 $, we obtain $\left(F^2 \right)_{\omega = 0} = 2 \zeta_r
\overline{T}$.  This yields
\be  D_R  =  \frac{ T_{{\cal A}}T_{{\cal B}}}{\zeta_{{\cal A}} T_{{\cal B}}
+\zeta_{{\cal B}} T_{{\cal A}}} +  \frac{1}{2} \frac{\left( T_{{\cal A}} -
T_{{\cal B}}\right)^2}{\zeta_{{\cal A}} T_{{\cal B}} + \zeta_{{\cal B}} T_{{\cal
A}}} \frac{\zeta_{{\cal A}}\zeta_{{\cal B}}}{\left( \zeta_{{\cal A}} +
\zeta_{{\cal B}} \right)^2} \ .  \ee
We see that the difference in temperatures, independently of the sign, enhances
the joint diffusion.

It is also instructive to write the Fokker-Planck equation in terms of the variables $r$ and $R$:
\be \begin{split} \partial_t P = & -  \left[ \frac{\zeta_{{\cal A}}+\zeta_{{\cal
B}}}{\zeta_{{\cal A}}\zeta_{{\cal B}}} \right] \partial_r \left( F(r) P \right) -
\\ &  - \left[ \frac{T_{{\cal B}} - T_{{\cal A}}}{\zeta_{{\cal A}} T_{{\cal B}}
+\zeta_{{\cal B}} T_{{\cal A}} }\right] F(r) \partial_R P + \\ & +  \left[
\frac{\zeta_{{\cal A}} T_{{\cal B}} +\zeta_{{\cal B}} T_{{\cal A}}}{\zeta_{{\cal
A}}\zeta_{{\cal B}}} \right] \partial^2_r P  \\ & +    \left[ \frac{ T_{{\cal A}}
T_{{\cal B}}}{\zeta_{{\cal A}} T_{{\cal B}} +\zeta_{{\cal B}} T_{{\cal A}}}
\right] \partial^2_R P  \ . \label{eq:FP_new}  \end{split} \ee
Here, the right hand side has explicitly the form of a divergence, and can be
written as
\be \partial_t P = - \partial_r J_r - \partial_R J_R \ , \ee
where the components of the flux are
\begin{subequations} \begin{align} J_r & = \left[ \frac{\zeta_{{\cal
A}}+\zeta_{{\cal B}}}{\zeta_{{\cal A}}\zeta_{{\cal B}}} \right] F(r) P - \left[
\frac{\zeta_{{\cal A}} T_{{\cal B}} +\zeta_{{\cal B}} T_{{\cal A}}}{\zeta_{{\cal
A}}\zeta_{{\cal B}}} \right] \partial_r P \ , \label{eq:flux_r}  \\
J_R & =  \left[  \frac{T_{{\cal B}} - T_{{\cal A}}}{\zeta_{{\cal A}} T_{{\cal B}}
+\zeta_{{\cal B}} T_{{\cal A}} } \right] F(r) P -  \left[ \frac{ T_{{\cal A}}
T_{{\cal B}}}{\zeta_{{\cal A}} T_{{\cal B}} +\zeta_{{\cal B}} T_{{\cal A}}}
\right] \partial_R P \ . \label{eq:flux_R} \end{align} \end{subequations}

\section{Hierarchy of equations for the correlation functions}

\subsection{Notations, definitions and symmetries}

As stated in the main text, we operate with a multidimensional Fokker-Planck
equation for the probability density $P$ as a function of the positions of all the
particles in the system.  More specifically, consider a system of $N_{\cal A}$
particles ${\cal A}$ and $N_{\cal B}$ particles ${\cal B}$; their coordinates
are $\mathbf{r}_1^{{\cal A}}, \mathbf{r}_2^{{\cal A}}, \ldots ,
\mathbf{r}_{N_{\cal A}}^{{\cal A}}$ and $\mathbf{r}_1^{{\cal B}},
\mathbf{r}_2^{{\cal B}}, \ldots , \mathbf{r}_{N_{\cal B}}^{{\cal B}}$. The
potential
energy includes single particle potentials and pairwise additive interactions:
\bea && U \left( \mathbf{r}_1^{{\cal A}}, \mathbf{r}_2^{{\cal A}}, \ldots ,
\mathbf{r}_{N_{\cal A}}^{{\cal A}} ; \mathbf{r}_1^{{\cal B}}, \mathbf{r}_2^{{\cal
B}}, \ldots , \mathbf{r}_{N_{\cal B}}^{{\cal B}} \right) = \nonumber \\ &  & \ \ \
\ \ \ \ \ \ \ \ \ \ \ \ \ \ = \sum_{i}^{N_{{\cal A}}} u_1^{{\cal A}} \left(
\mathbf{r}_i^{{\cal A}}\right) + \sum_{j}^{N_{{\cal B}}} u_1^{{\cal B}} \left(
\mathbf{r}_j^{{\cal A}} \right) + \nonumber \\ &  & \ \ \ \ \ \ \ \ \ \ \ \ \ \ \
\ \ + \frac{1}{2} \sum_{i\neq j}^{N_{{\cal A}}} u^{{\cal AA}} \left(
\mathbf{r}_i^{{\cal A}} - \mathbf{r}_j^{{\cal A}} \right) + \nonumber \\ &  &  \ \
\ \ \ \ \ \ \ \ \ \ \ \ \ \ \ + \sum_{i}^{N_{{\cal A}}} \sum_{j}^{N_{{\cal B}}}
u^{{\cal AB}} \left( \mathbf{r}_i^{{\cal A}} - \mathbf{r}_j^{{\cal B}} \right) +
\nonumber \\ &  & \ \ \ \ \ \ \ \ \ \ \ \ \ \ \ \ \ + \frac{1}{2} \sum_{i\neq
j}^{N_{{\cal B}}} u^{{\cal BB}} \left( \mathbf{r}_i^{{\cal B}} -
\mathbf{r}_j^{{\cal B}} \right) \ . \label{eq:pairwise_additive_interactions} \eea
The probability density $P$ is also a function of all the coordinates: $P =
P\left( \mathbf{r}_1^{{\cal A}}, \mathbf{r}_2^{{\cal A}}, \ldots ,
\mathbf{r}_{N_{\cal A}}^{{\cal A}} ; \mathbf{r}_1^{{\cal B}}, \mathbf{r}_2^{{\cal
B}}, \ldots , \mathbf{r}_{N_{\cal B}}^{{\cal B}} \right)$. It is normalized:
\be \int P\left( \mathbf{r}_1^{{\cal A}}, \mathbf{r}_2^{{\cal A}}, \ldots ,
\mathbf{r}_{N_{\cal A}}^{{\cal A}} ; \mathbf{r}_1^{{\cal B}}, \mathbf{r}_2^{{\cal
B}}, \ldots , \mathbf{r}_{N_{\cal B}}^{{\cal B}} \right) d \{ \mathbf{r} \} = 1 \
. \ee
Define single particle probability densities, two particle probability densities,
etc as:
\bseq p_1^{{\cal A}}(\mathbf{r}) & =  \int \delta \left( \mathbf{r}_{i}^{{\cal A}}
- \mathbf{r} \right) P \left( \{ \mathbf{r} \} \right) d \{ \mathbf{r} \} \\
p_1^{{\cal B}}(\mathbf{r}) & =  \int \delta \left( \mathbf{r}_{i}^{{\cal B}} -
\mathbf{r} \right) P \left( \{ \mathbf{r} \} \right) d \{ \mathbf{r} \}  \\
p_2^{{\cal AA}}(\mathbf{r}, \mathbf{r}^{\prime} ) & =  \int \delta \left(
\mathbf{r}_i^{{\cal A}} - \mathbf{r} \right) \delta \left( \mathbf{r}_{j}^{{\cal
A}} - \mathbf{r}^{\prime} \right) P \left( \{ \mathbf{r} \} \right) d \{
\mathbf{r} \} \\ p_2^{{\cal BB}}(\mathbf{r}, \mathbf{r}^{\prime} ) & =  \int
\delta \left( \mathbf{r}_i^{{\cal B}} - \mathbf{r} \right) \delta \left(
\mathbf{r}_{j}^{{\cal B}} - \mathbf{r}^{\prime} \right) P \left( \{ \mathbf{r} \}
\right) d \{ \mathbf{r} \} \\ p_2^{{\cal AB}}(\mathbf{r}, \mathbf{r}^{\prime} ) &
=  \int \delta \left( \mathbf{r}_i^{{\cal A}} - \mathbf{r} \right) \delta \left(
\mathbf{r}_{j}^{{\cal B}} - \mathbf{r}^{\prime} \right) P \left( \{ \mathbf{r} \}
\right) d \{ \mathbf{r} \} \end{align} \end{subequations}
There are 4 types of 3-particle densities: $p_3^{{AAA}}(\mathbf{r},
\mathbf{r}^{\prime}, \mathbf{r}^{\prime\prime} )$, $p_3^{{AAB}}(\mathbf{r},
\mathbf{r}^{\prime}, \mathbf{r}^{\prime\prime} )$, $p_3^{{ABB}}(\mathbf{r},
\mathbf{r}^{\prime}, \mathbf{r}^{\prime\prime} )$, $p_3^{{BBB}}(\mathbf{r},
\mathbf{r}^{\prime}, \mathbf{r}^{\prime\prime} )$.  All these densities are
independent of $i$ and $j$ etc, i.e., the probability density is the same for
every particle of a given species.

There are several normalization and symmetry properties (${{\cal X},{\cal Y} = {\cal A},{\cal B}}$):
\begin{subequations} \begin{align} & \int p_1^{{\cal X}} (\mathbf{r} ) d\mathbf{r} = 1 \ , \\ & p_2^{{\cal X}{\cal X}} (\mathbf{r},\mathbf{r}^{\prime} )  =
p_2^{{\cal X}{\cal X}} (\mathbf{r}^{\prime},\mathbf{r} ) \\ & p_2^{{\cal XY}}
(\mathbf{r},\mathbf{r}^{\prime} )  = p_2^{{\cal Y}{\cal X}} (\mathbf{r}^{\prime},
\mathbf{r} ) \ , \\ & \int p_2^{{\cal XY}} (\mathbf{r},\mathbf{r}^{\prime} )
d\mathbf{r} d\mathbf{r}^{\prime}  = 1 \ , \\ & \int p_2^{{\cal XY}} (\mathbf{r},
\mathbf{r}^{\prime} ) d\mathbf{r}^{\prime}  = p_1^{{\cal X}} (\mathbf{r})  \ , \\
& \int p_2^{{\cal XY}} (\mathbf{r},\mathbf{r}^{\prime} ) d\mathbf{r}  = p_1^{{\cal
Y}} (\mathbf{r}^{\prime}) \end{align} \end{subequations}

\subsection{Fokker-Planck equations for the densities}

Integrating out all variables except for one, or except for two, etc, we obtain the following dynamic equations for the densities:

\be \begin{split} \frac{ \partial p_1^{{\cal A}}(\mathbf{r})}{\partial t} & =
\frac{T_{{\cal A}}}{\zeta_{{\cal A}}} \nabla^2_{\mathbf{r}} p_1^{{\cal A}}
(\mathbf{r}) +   \frac{1}{\zeta_{{\cal A}}} \partial_{\mathbf{r}} \left[ \frac{
\partial u_1^{{\cal A}}(\mathbf{r})}{\partial \mathbf{r}} p_1^{{\cal A}}
(\mathbf{r}) \right] + \\ & +    \frac{N_{{\cal A}}-1}{\zeta_{{\cal A}}}
\partial_{\mathbf{r}} \left[ \int \frac{\partial u^{{\cal AA}}\left( \mathbf{r},
\mathbf{r}^{\prime} \right)}{{\partial \mathbf{r}}}  p_2^{{\cal AA}} \left(
\mathbf{r}, \mathbf{r}^{\prime} \right) d \mathbf{r}^{\prime} \right] + \\ & +
\frac{N_{{\cal B}}}{\zeta_{{\cal B}}} \partial_{\mathbf{r}} \left[ \int
\frac{\partial u^{{\cal AB}} \left( \mathbf{r}, \mathbf{r}^{\prime} \right)}
{\partial {\mathbf{r}}} p_2^{{\cal AB}} \left( \mathbf{r}, \mathbf{r}^{\prime}
\right) d \mathbf{r}^{\prime} \right] \ . \label{eq:Equation_for _p1} \end{split}
\ee

A similar equation is obtained for $p_1^{{\cal B}}(\mathbf{r})$, which is not
given for brevity. For large numbers of particles, we can replace $N_{{\cal A}} -
1 \simeq N_{{\cal A}}$.
\be \begin{split} & \frac{\partial p_2^{{\cal AA}}(\mathbf{r},
\mathbf{r}^{\prime})}{\partial t}  =  \frac{T_{{\cal A}}}{\zeta_{{\cal A}}} \left(
\nabla^2_{\mathbf{r}} +  \nabla^2_{\mathbf{r}^{\prime}} \right) p_2^{{\cal AA}}
(\mathbf{r}, \mathbf{r}^{\prime}) + \\ & +  \frac{1}{\zeta_{{\cal A}}}
\partial_{\mathbf{r}} \left[ \frac{ \partial u_1^{{\cal A}}(\mathbf{r}) }{\partial
\mathbf{r}} p_2^{{\cal AA}}(\mathbf{r}, \mathbf{r}^{\prime}) \right] +  \\ & +
\frac{1}{\zeta_{{\cal A}}} \partial_{\mathbf{r}} \left[ \frac{ \partial u^{{\cal
AA}}\left(\mathbf{r}, \mathbf{r}^{\prime} \right) }{\partial \mathbf{r}}
p_2^{{\cal AA}}(\mathbf{r}, \mathbf{r}^{\prime}) \right] + \\ & +  \frac{N_{{\cal
A}}}{\zeta_{{\cal A}}} \partial_{\mathbf{r}} \left[ \int \frac{ \partial u^{{\cal
AA}}\left(\mathbf{r}, \mathbf{r}^{\prime\prime} \right) }{\partial \mathbf{r}}
p_3^{{AAA}}(\mathbf{r}, \mathbf{r}^{\prime},\mathbf{r}^{\prime\prime})
d\mathbf{r}^{\prime\prime}\right] + \\ & +  \frac{N_{{\cal B}}}{\zeta_{{\cal A}}}
\partial_{\mathbf{r}} \left[ \int \frac{ \partial u^{{\cal AB}}\left(\mathbf{r},
\mathbf{r}^{\prime\prime} \right) }{\partial \mathbf{r}} p_3^{{AAB}}(\mathbf{r},
\mathbf{r}^{\prime},\mathbf{r}^{\prime\prime}) d\mathbf{r}^{\prime\prime}\right] +
\\ & +  \frac{1}{\zeta_{{\cal A}}} \partial_{\mathbf{r}^{\prime}} \left[ \frac{
\partial u_1^{{\cal A}}(\mathbf{r}^{\prime}) }{\partial \mathbf{r}^{\prime}}
p_2^{{\cal AA}}(\mathbf{r}, \mathbf{r}^{\prime}) \right] + \\ & +  \frac{1}
{\zeta_{{\cal A}}} \partial_{\mathbf{r}^{\prime}} \left[ \frac{ \partial u^{{\cal
AA}}\left(\mathbf{r}, \mathbf{r}^{\prime} \right) }{\partial \mathbf{r}^{\prime}}
p_2^{{\cal AA}}(\mathbf{r}, \mathbf{r}^{\prime}) \right] + \\ & +  \frac{N_{{\cal
A}}}{\zeta_{{\cal A}}} \partial_{\mathbf{r}^{\prime}} \left[ \int \frac{ \partial
u^{{\cal AA}}\left(\mathbf{r}^{\prime}, \mathbf{r}^{\prime\prime} \right) }
{\partial \mathbf{r}^{\prime}} p_3^{{AAA}}(\mathbf{r}, \mathbf{r}^{\prime},
\mathbf{r}^{\prime\prime}) d\mathbf{r}^{\prime\prime}\right] +  \\ & +
\frac{N_{{\cal B}}}{\zeta_{{\cal A}}} \partial_{\mathbf{r}^{\prime}} \left[ \int
\frac{ \partial u^{{\cal AB}}\left(\mathbf{r}^{\prime}, \mathbf{r}^{\prime\prime}
\right) }{\partial \mathbf{r}^{\prime}} p_3^{{AAB}}(\mathbf{r},
\mathbf{r}^{\prime},\mathbf{r}^{\prime\prime}) d\mathbf{r}^{\prime\prime}\right]
\ . \label{eq:Equation_for _p2AA} \end{split}\ee
A similar equation (not given here) is obtained for $p_2^{{\cal BB}}$. But the
equation for the mixed probability deserves to be written down:
\be \begin{split} & \frac{\partial p_2^{{\cal AB}}(\mathbf{r},
\mathbf{r}^{\prime})}{\partial t}  =  \left( \frac{T_{{\cal A}}}{\zeta_{{\cal A}}}
\nabla^2_{\mathbf{r}} +  \frac{T_{{\cal B}}}{\zeta_{{\cal B}}}
\nabla^2_{\mathbf{r}^{\prime}} \right) p_2^{{\cal AB}}(\mathbf{r},
\mathbf{r}^{\prime}) +  \\ & +  \frac{1}{\zeta_{{\cal A}}} \partial_{\mathbf{r}}
\left[ \frac{ \partial u_1^{{\cal A}}(\mathbf{r}) }{\partial \mathbf{r}}
p_2^{{\cal AB}}(\mathbf{r}, \mathbf{r}^{\prime}) \right] + \\ & +  \frac{1}
{\zeta_{{\cal A}}} \partial_{\mathbf{r}} \left[ \frac{ \partial u^{{\cal
AB}}\left(\mathbf{r}, \mathbf{r}^{\prime} \right) }{\partial \mathbf{r}}
p_2^{{\cal AB}}(\mathbf{r}, \mathbf{r}^{\prime}) \right] + \\ & +  \frac{N_{{\cal
A}}}{\zeta_{{\cal A}}} \partial_{\mathbf{r}} \left[ \int \frac{ \partial u^{{\cal
AA}}\left(\mathbf{r}, \mathbf{r}^{\prime\prime} \right) }{\partial \mathbf{r}}
p_3^{{ABA}}(\mathbf{r}, \mathbf{r}^{\prime},\mathbf{r}^{\prime\prime})
d\mathbf{r}^{\prime\prime}\right] + \\ & +  \frac{N_{{\cal B}}}{\zeta_{{\cal A}}}
\partial_{\mathbf{r}} \left[ \int \frac{ \partial u^{{\cal AB}}\left(\mathbf{r},
\mathbf{r}^{\prime\prime} \right) }{\partial \mathbf{r}} p_3^{{ABB}}(\mathbf{r},
\mathbf{r}^{\prime},\mathbf{r}^{\prime\prime}) d\mathbf{r}^{\prime\prime}\right] +
\\ & +  \frac{1}{\zeta_{{\cal B}}} \partial_{\mathbf{r}^{\prime}} \left[ \frac{
\partial u_1^{{\cal B}}(\mathbf{r}^{\prime}) }{\partial \mathbf{r}^{\prime}}
p_2^{{\cal AB}}(\mathbf{r}, \mathbf{r}^{\prime}) \right] + \\ & +  \frac{1}
{\zeta_{{\cal B}}} \partial_{\mathbf{r}^{\prime}} \left[ \frac{ \partial u^{{\cal
AB}}\left(\mathbf{r}, \mathbf{r}^{\prime} \right) }{\partial \mathbf{r}^{\prime}}
p_2^{{\cal AB}}(\mathbf{r}, \mathbf{r}^{\prime}) \right] +  \\ & +  \frac{N_{{\cal
A}}}{\zeta_{{\cal B}}} \partial_{\mathbf{r}^{\prime}} \left[ \int \frac{ \partial
u^{{BA}}\left(\mathbf{r}^{\prime}, \mathbf{r}^{\prime\prime} \right) }{\partial
\mathbf{r}^{\prime}} p_3^{{ABA}}(\mathbf{r}, \mathbf{r}^{\prime},
\mathbf{r}^{\prime\prime}) d\mathbf{r}^{\prime\prime}\right] +  \\ & +
\frac{N_{{\cal B}}}{\zeta_{{\cal B}}} \partial_{\mathbf{r}^{\prime}} \left[ \int
\frac{ \partial u^{{\cal BB}}\left(\mathbf{r}^{\prime}, \mathbf{r}^{\prime\prime}
\right) }{\partial \mathbf{r}^{\prime}} p_3^{{ABB}}(\mathbf{r},
\mathbf{r}^{\prime},\mathbf{r}^{\prime\prime}) d\mathbf{r}^{\prime\prime}\right]
\ . \label{eq:Equation_for _p2AB} \end{split} \ee
The equations for $p_3$ involve $p_4$, and so on, \textit{ad infinitum}.

If the density is small enough, we can neglect all triple collisions, i.e.,
directly discard all terms involving $p_3$.  Indeed, given the normalization, any
term containing $p_2$ in equations (\ref{eq:Equation_for _p2AA}) or
(\ref{eq:Equation_for _p2AA}) is of order $1/V^2$, while every term containing
$p_3$ is of order $N/V^3$. This, of course, simplifies the equations quite
dramatically, and
reduces them essentially to what we obtained for the case of two particles.
Omitting the single particle potential terms, we obtain
\be \begin{split} \frac{\partial p_2^{{\cal AA}}(\mathbf{r},\mathbf{r}^{\prime})}
{\partial t}  & =   \frac{T_{{\cal A}}}{\zeta_{{\cal A}}} \left(
\nabla^2_{\mathbf{r}} +  \nabla^2_{\mathbf{r}^{\prime}} \right) p_2^{{\cal AA}}
(\mathbf{r}, \mathbf{r}^{\prime}) + \\ & +  \frac{1}{\zeta_{{\cal A}}}
\partial_{\mathbf{r}} \left[ \frac{ \partial u^{{\cal AA}}\left(\mathbf{r},
\mathbf{r}^{\prime} \right) }{\partial \mathbf{r}} p_2^{{\cal AA}}(\mathbf{r},
\mathbf{r}^{\prime}) \right] + \\ & +  \frac{1}{\zeta_{{\cal A}}}
\partial_{\mathbf{r}^{\prime}} \left[ \frac{ \partial u^{{\cal
AA}}\left(\mathbf{r}, \mathbf{r}^{\prime} \right) }{\partial \mathbf{r}^{\prime}}
p_2^{{\cal AA}}(\mathbf{r}, \mathbf{r}^{\prime}) \right] \ ,
\label{eq:Equation_for _p2AA_without_p3} \end{split} \ee
(and a similar equation for $p_2^{{\cal BB}}$),
\be \begin{split} \frac{\partial p_2^{{\cal AB}}(\mathbf{r},\mathbf{r}^{\prime})}
{\partial t} & =   \left( \frac{T_{{\cal A}}}{\zeta_{{\cal A}}}
\nabla^2_{\mathbf{r}} +  \frac{T_{{\cal B}}}{\zeta_{{\cal B}}}
\nabla^2_{\mathbf{r}^{\prime}} \right) p_2^{{\cal AB}}(\mathbf{r},
\mathbf{r}^{\prime}) + \\  & +  \frac{1}{\zeta_{{\cal A}}} \partial_{\mathbf{r}}
\left[ \frac{ \partial u^{{\cal AB}}\left(\mathbf{r}, \mathbf{r}^{\prime} \right)
}{\partial \mathbf{r}} p_2^{{\cal AB}}(\mathbf{r}, \mathbf{r}^{\prime}) \right] +
\\  & +  \frac{1}{\zeta_{{\cal B}}} \partial_{\mathbf{r}^{\prime}} \left[ \frac{
\partial u^{{\cal AB}}\left(\mathbf{r}, \mathbf{r}^{\prime} \right) }{\partial
\mathbf{r}^{\prime}} p_2^{{\cal AB}}(\mathbf{r}, \mathbf{r}^{\prime}) \right]   \
. \label{eq:Equation_for _p2AB_without_p3} \end{split} \ee
These equations are simple enough to guess the solution based on our knowledge of the two-particle case:
\begin{subequations} \label{eq:ansatz_for_p2} \begin{align} p_2^{{\cal AA}}
(\mathbf{r},\mathbf{r}^{\prime}) & = p_1^{{\cal A}}(\mathbf{r}) p_1^{{\cal A}}
(\mathbf{r}^{\prime}) \exp \left[ - \frac{u^{{\cal AA}} \left( \mathbf{r} -
\mathbf{r}^{\prime} \right)}{T_{{\cal A}}} \right] \\ p_2^{{\cal BB}}(\mathbf{r},
\mathbf{r}^{\prime}) & = p_1^{{\cal B}}(\mathbf{r}) p_1^{{\cal B}}
(\mathbf{r}^{\prime}) \exp \left[ - \frac{u^{{\cal BB}} \left( \mathbf{r} -
\mathbf{r}^{\prime} \right)}{T_{{\cal B}}} \right] \\ p_2^{{\cal AB}}(\mathbf{r},
\mathbf{r}^{\prime}) & = p_1^{{\cal A}}(\mathbf{r}) p_1^{{\cal B}}
(\mathbf{r}^{\prime}) \exp \left[ - \frac{u^{{\cal AB}} \left( \mathbf{r} -
\mathbf{r}^{\prime} \right)}{\overline{T}} \right] \end{align} \end{subequations}
Of course, the central feature of this result is the appearance of the average temperature, as defined in the main text
\be \overline{T} = \frac{\zeta_{{\cal A}} T_{{\cal B}} + \zeta_{{\cal B}} T_{{\cal A}}}{\zeta_{{\cal A}} + \zeta_{{\cal B}}} \ , \ee.

In order to obtain the result (\ref{eq:ansatz_for_p2}), we look for a solution of
the form $p_2^{{\cal XY}}(\mathbf{r},\mathbf{r}^{\prime}) = q^{{\cal XY}}
(\mathbf{r},\mathbf{r}^{\prime}) \exp \left[ - \beta u^{{\cal XY}}(\mathbf{r},
\mathbf{r}^{\prime}) \right]$, plug it into equations (\ref{eq:Equation_for
_p2AA_without_p3}) or (\ref{eq:Equation_for _p2AB_without_p3}), and discover, that
in the remaining equation for $q$, the variables separate, meaning that $q$ factorizes into
a factor that depends only on $\mathbf{r}$ and a factor that depends only on
$\mathbf{r}^{\prime}$.

\subsection{Diffusion equations and non-equilibrium chemical potentials}

Plugging the \textit{ansatz} (\ref{eq:ansatz_for_p2}) into equations (\ref{eq:Equation_for _p1}), we obtain closed results for the densities.  This involves the integral
\be\int \frac{\partial u^{{\cal AA}}\left( \mathbf{r}, \mathbf{r}^{\prime} \right)}{{\partial \mathbf{r}}} p_1^{{\cal A}} \left( \mathbf{r}^{\prime} \right) e^{-\frac{u^{{\cal AA}}\left( \mathbf{r}, \mathbf{r}^{\prime} \right)}{T_{{\cal A}}}} d \mathbf{r}^{\prime} \ , \ee
which can be integrated by parts.
Finally, we obtain equations that look like diffusion equations for a regular system in contact with a thermostat,
\begin{subequations} \label{eq:Equations_for _p1_closed} \begin{align} \frac{
\partial c^{{\cal A}}(\mathbf{r})}{\partial t} & =  \frac{1}{\zeta_{{\cal A}}}
\frac{\partial}{\partial \mathbf{r}} \left( c^{{\cal A}} \frac{\partial \mu_{{\cal
A}} }{\partial {\mathbf{r}}}\right) \\ \frac{ \partial c^{{\cal B}}(\mathbf{r})}
{\partial t} & =  \frac{1}{\zeta_{{\cal B}}} \frac{\partial}{\partial
{\mathbf{r}}} \left( c^{{\cal B}} \frac{\partial \mu_{{\cal B}}}{\partial
{\mathbf{r}}} \right)  \ , \end{align} \end{subequations}
but these equations contain \textit{non-equilibrium chemical
potentials}, as stated in the main text (Eq.(11)).

\subsection{Linear stability analysis}

Suppose that $c_0^{{\cal A}}$ and $c_0^{{\cal B}}$ are the averaged spatially uniform concentrations of both components.  By introducing small space dependent perturbations $c^{{\cal A}}(\mathbf{r}) = c_0^{{\cal A}} + \delta c^{{\cal A}}
(\mathbf{r})$ and $c^{{\cal B}}(\mathbf{r}) = c_0^{{\cal B}} + \delta c^{{\cal B}}(\mathbf{r})$, we perform a linear stability analysis in the standard way:
\be \left\{ \begin{array}{l} \frac{\partial \delta c^{{\cal A}}}{\partial t}  =
\frac{1}{\zeta_{{\cal A}}} \nabla^2 \left[ \left( T_{{\cal A}} + T_{{\cal A}}
c_0^{{\cal A}} B_{{\cal A}} \right) \delta c^{{\cal A}} + \left( \overline{T}
c_0^{{\cal A}} B_{{\cal AB}} \right) \delta c^{{\cal B}}  \right] \\ \\
\frac{\partial \delta c^{{\cal B}}}{\partial t}  = \frac{1}{\zeta_{{\cal B}}}
\nabla^2 \left[\left( \overline{T} c_0^{{\cal B}} B_{{\cal AB}} \right) \delta
c^{{\cal A}} + \left( T_{{\cal B}} + T_{{\cal B}} c_0^{{\cal B}} B_{{\cal B}}
\right) \delta c^{{\cal B}} \right] \end{array}\right. \ee
This shows that an instability occurs (at $q=0$, i.e., macroscopically) under the condition that the determinant of this matrix vanishes, i.e., the system is unstable if
\be c_0^{\cal A} c_0^{\cal B} \overline{T}^2 B_{{\cal AB}}^2 > T_AT_B (1 +
c_0^{\cal A} B_{{\cal A}} )(1+ c_0^{\cal B} B_{{\cal B}}) \ .
\label{eq:general_spinodal_condition} \ee

At the instability, the unstable combination (eigenvector whose eigenvalue flips sign) is
\be \frac{\delta c^{{\cal A}} (\mathbf{r}) \zeta_{{\cal A}}}{\sqrt{ T_{{\cal A}}
B_{{\cal A}} c_0^{{\cal A}} \left( 1 + B_{{\cal A}} c_0^{{\cal A}} \right)}} -
\frac{\delta c^{{\cal B}} (\mathbf{r}) \zeta_{{\cal B}}}{\sqrt{ T_{{\cal B}}
B_{{\cal B}} c_0^{{\cal B}} \left( 1 + B_{{\cal B}} c_0^{{\cal B}} \right)}} \ .
\ee

\begin{figure}
  \centering
  \includegraphics[width=0.3\textwidth]{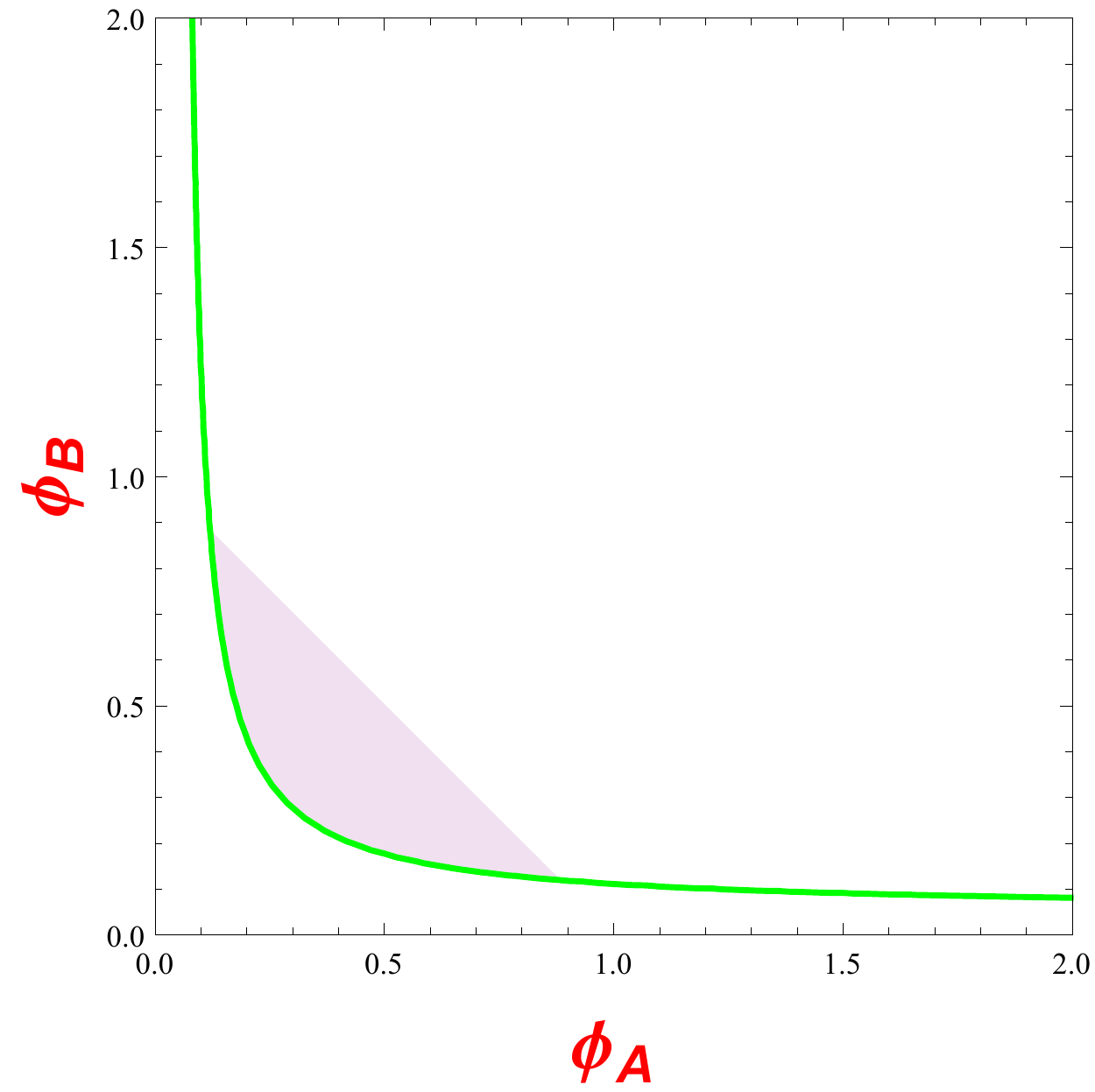}\\
  \caption{Spinodal line (\ref{eq:general_spinodal_condition}) (in green), shown
  for one particular choice of parameters ($T_{{\cal A}}/T_{{\cal B}},
  \zeta_{{\cal A}}/\zeta_{{\cal B}}, B_{{\cal A}}B_{{\cal B}}/B_{{\cal AB}}^2$).
  Physical meaning has only region $\phi_{{\cal A}} + \phi_{{\cal B}} <1$, the
  instability region within this region is shaded.  }\label{fig:Hyperbola}
\end{figure}

In the plane ($c_0^{{\cal A}}$, $c_0^{{\cal B}}$), the
spinodal line (\ref{eq:general_spinodal_condition}) is a hyperbola (Fig.
\ref{fig:Hyperbola}).  A better way to represent it is to use a triangular phase diagram as given in Fig.2 of the main text.

\section{Non-equilibrium ``spinodal'' line for athermal particles}

The dimensionless parameters of the system are as follows:
\begin{itemize} \item Contrast of excluded volumes,
\be \beta = \frac{B_{{\cal AB}}}{\sqrt{B_{{\cal A}} B_{{\cal B}}}} \ . \ee
\item Contrast of temperatures,
\be \tau = \frac{T_{{\cal A}}-T_{{\cal B}}}{T_{{\cal A}}+T_{{\cal B}}} \ ; \ \ -1 < \tau < 1 \ . \ee
\item Contrast of frictions,
\be \kappa = \frac{\zeta_{{\cal A}}-\zeta_{{\cal B}}}{\zeta_{{\cal A}}+\zeta_{{\cal B}}} \ ; \ \ -1 < \kappa < 1 \ . \ee
\end{itemize}

\begin{figure}
  \centering
  \includegraphics[width=0.45\textwidth]{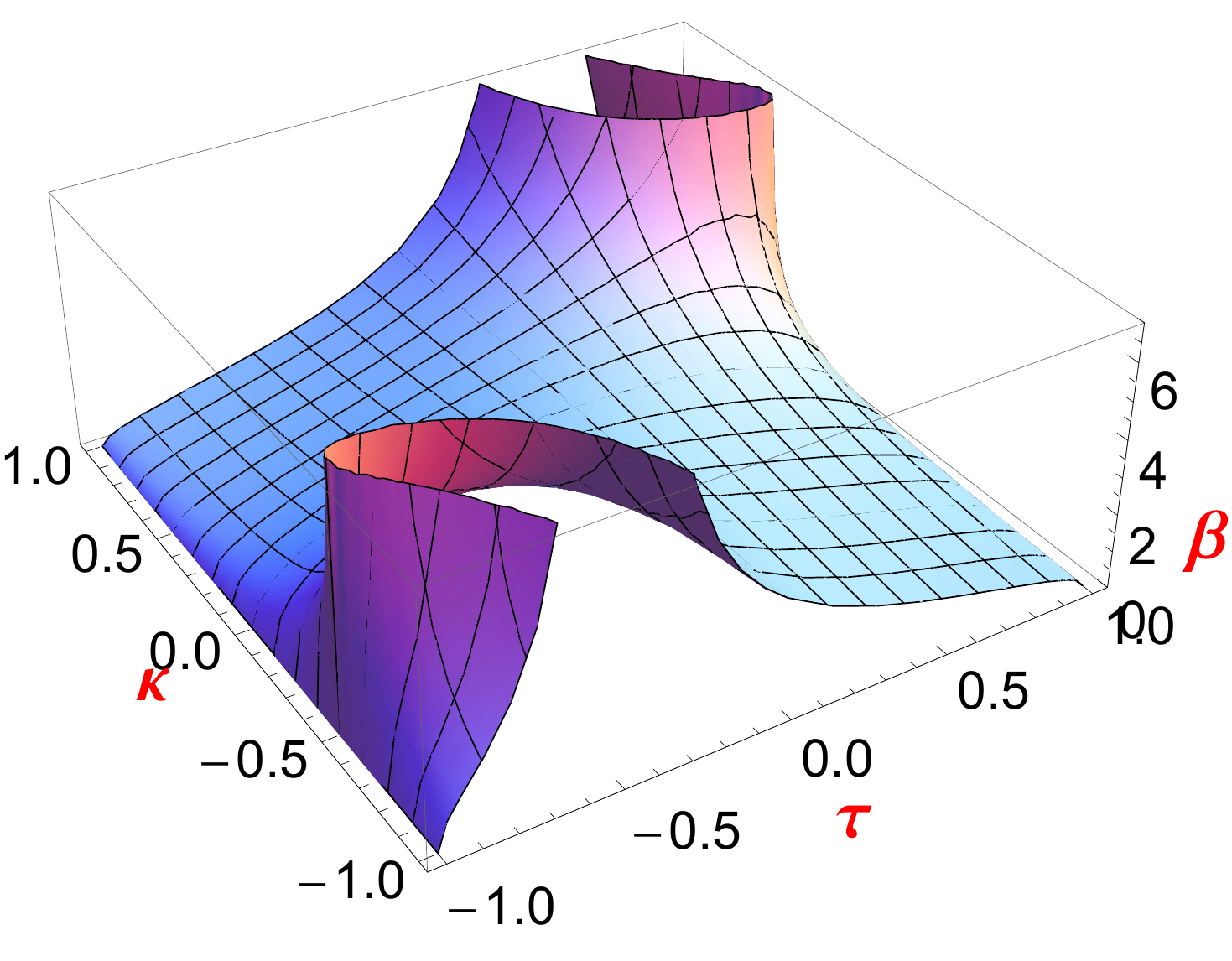}\\
  \caption{Instability exists above this surface, see condition (\ref{eq:beta_kappa_tau_condition}).}\label{fig:Kappa_Tau_Beta_3D}
\end{figure}

\begin{figure}
  \centering
  \includegraphics[width=0.45\textwidth]{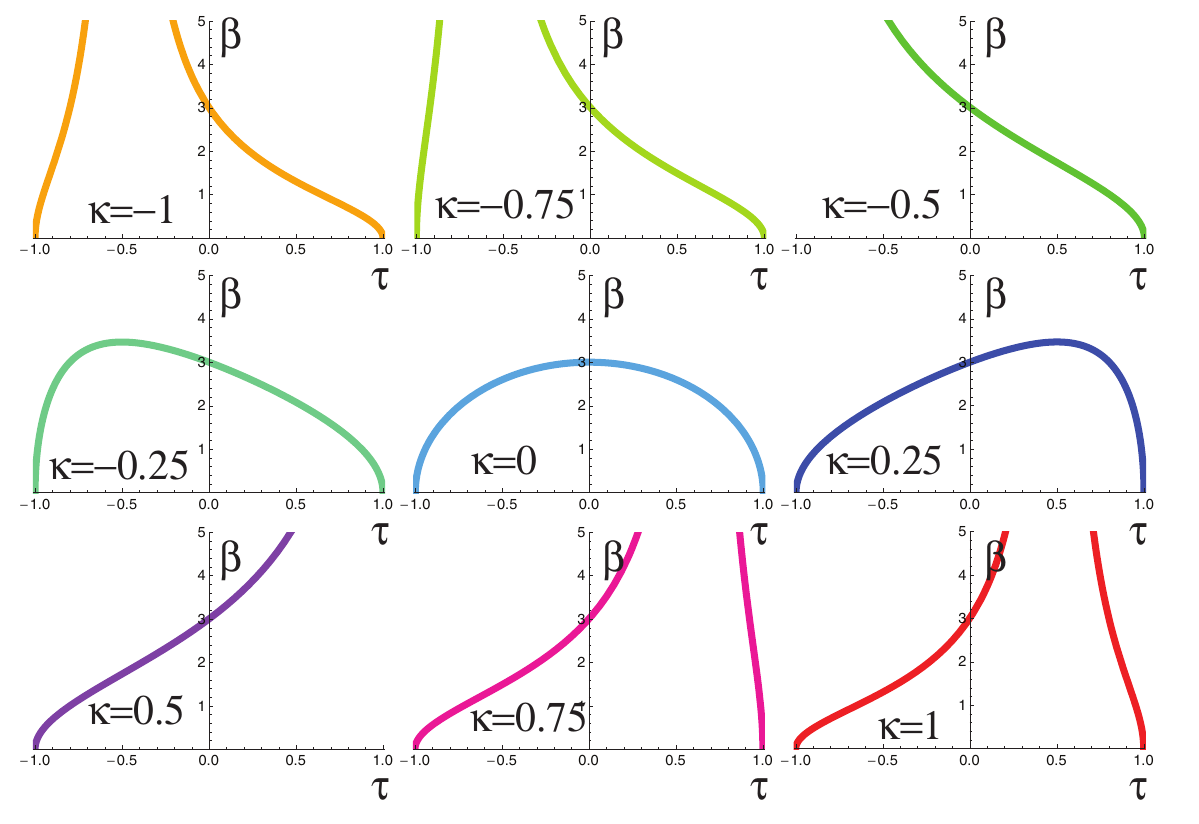}\\
  \caption{For every $\kappa$, instability exists above the line, according to Eqns (\ref{eq:beta_kappa_tau_condition}).}\label{fig:Kappa_Tau_Beta}
\end{figure}

The condition that the spinodal exists within the physical range $\phi_{{\cal A}} + \phi_{{\cal B}} <1$ reads
\be  \frac{4 \kappa^2  \tau^2  - 4 \kappa \tau +1}{1-\tau^2} \beta^2 > 9 \ . \label{eq:beta_kappa_tau_condition} \ee
This condition is presented graphically in two different ways, in Fig.
\ref{fig:Kappa_Tau_Beta_3D} in the form of a 3D surface and in Fig.
\ref{fig:Kappa_Tau_Beta} as an array of 2D plots.  Beautifully, the contrast of
frictions $\kappa$ becomes irrelevant for the equilibrium system, when $\tau = 0$.

\section{Non-equilibrium osmotic pressure and ``binodal''}

To address the steady state phase segregation, in addition to the non-equilibrium
chemical potentials we also need to define a non-quilibrium osmotic pressure.  We
derive it in two different ways.

\subsection{Derivation 1}

\begin{figure}
  \centering
  \includegraphics[width=0.25\textwidth]{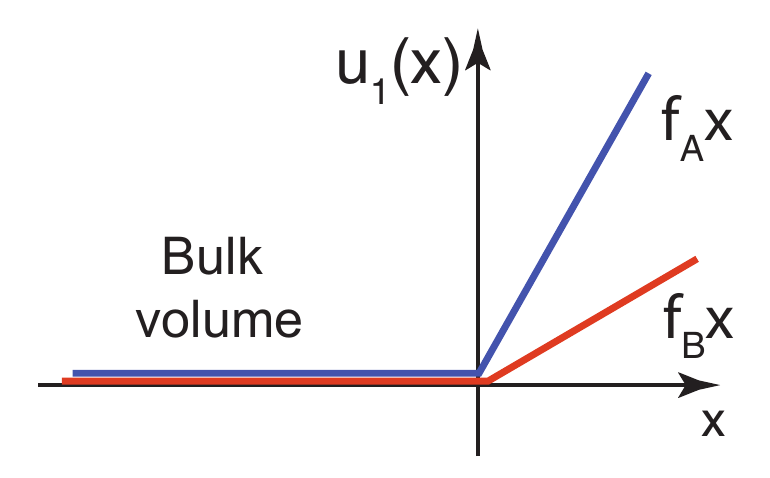}\\
  \caption{Ramp potential used to calculate pressure.}\label{fig:Ramp_potential}
\end{figure}

To find the osmotic pressure, imagine that the system ``feels'' single particle
potentials $u_1^{{\cal A}}(\mathbf{r})$ and $u_1^{{\cal B}}(\mathbf{r})$ such that
they are both like a box, except that one wall of this box has a (not necessarily very)
sharp potential ``ramp'' in the direction, perpendicular to the wall: $u_1^{{\cal
A,B}}(\mathbf{r}) = f_{{\cal A,B}} x$, as shown in Fig.\ref{fig:Ramp_potential}.  In this case, the pressure is found according to
\be p = f_{{\cal A}} \int_{0}^{\infty} c^{{\cal A}}(x) dx +f_{{\cal B}} \int_{0}^{\infty} c^{{\cal B}}(x) dx \ , \label{eq:compute_pressure} \ee
because every particle ${{\cal A}}$ present in the ramp area exerts on the wall the force $f_{{\cal A}}$, and similarly for ${{\cal B}}$.  We emphasize, that this
is actually an \textit{osmotic} pressure, in the sense that the ramp potentials
$u_1^{{\cal A,B}}(\mathbf{r}) = f_{{\cal A,B}} x$ act only on the ${{\cal A}}$ and
${{\cal B}}$ particles while the solvent penetrates everywhere completely freely.
This means that our ramp potentials represent an osmotic piston.

A similar expression for the osmotic pressure was also used in Ref.
\cite{Many_authors_pressure_2014_1}, where it is derived from the expression of
the Helmhotz partition sum, i.e., from equilibrium statistical mechanics.  We feel
necessary to emphasize that Eq. (\ref{eq:compute_pressure}) is derived on purely
mechanical grounds, and it has nothing to do with thermodynamic equilibrium.  As
such, it is perfectly applicable to our present problem.

To find the steady state concentration profile in the presence of ramp potentials,
we slightly generalize the diffusion equations (\ref{eq:Equations_for _p1_closed})
by including the external potentials $u_1$:
\be \mu_{{\cal A}} \to \mu_{{\cal A}} + u_1^{{\cal A}} \ \mathrm{and} \ \mu_{{\cal
B}} \to \mu_{{\cal B}} + u_1^{{\cal B}} \ . \ee
At steady state, the concentration profile must be such that $\mu + u_1 =
\mathrm{const}$ for both the ${{\cal A}}$ and ${{\cal B}}$ components. This can be
written as
\begin{subequations} \begin{align} c^{{\cal A}}(\mathbf{r})  & =  C^{{\cal A}}
e^{- \frac{u_1^{{\cal A}}}{T_{{\cal A}}}} \left[ 1  +  c^{{\cal A}}(\mathbf{r})
B_{{\cal A}}  +  c^{{\cal B}}(\mathbf{r}) B_{{\cal AB}} \frac{\overline{T}}
{T_{{\cal A}}} \right] \\ c^{{\cal B}}(\mathbf{r})  & =  C^{{\cal B}} e^{-
\frac{u_1^{{\cal B}}}{T_{{\cal B}}}} \left[ 1  +  c^{{\cal B}}(\mathbf{r})
B_{{\cal B}}  +  c^{{\cal A}}(\mathbf{r}) B_{{\cal AB}} \frac{\overline{T}}{T_{{\cal B}}} \right] \end{align}\end{subequations}
Here $C^{{\cal A}}$ and $C^{{\cal B}}$ are normalization factors.  To make things
simple, we assume that the ``ramps'' are not too shallow, such that
the normalization integral is dominated by the bulk volume $V$ where both ramp
potentials vanish. Given that the virial terms in the chemical potentials are the
corrections to the ideal gas, we solve iteratively and get
\be \begin{split} c^{{\cal A}}(\mathbf{r}) & =  \frac{N_{{\cal A}}}{V} e^{-
\frac{u_1^{{\cal A}}}{T_{{\cal A}}}} \left[ 1  +  \frac{N_{{\cal A}}}{V} B_{{\cal
A}} \left( 1- e^{- \frac{u_1^{{\cal A}}}{T_{{\cal A}}}} \right) + \right. \\ & +
\left.   \frac{N_{{\cal B}}}{V} B_{{\cal AB}} \frac{\overline{T}}{T_{{\cal A}}}
\left( 1- e^{- \frac{u_1^{{\cal B}}}{T_{{\cal B}}}} \right) \right] \end{split}
\ee
and similarly for $c^{{\cal B}}(\mathbf{r})$. Note that the result does not depend on the ramp forces $f_{{\cal
A}}$ and $f_{{\cal B}}$, which do not have to be identical.

\subsection{Derivation 2}

Our starting point of the second derivation is the kinetic expression of the
pressure
\be p = p_{\mathrm{ideal}} - \frac{N}{6V} \left< \sum_{i\neq j} \mathbf{r}_{ij} \cdot \partial_{\mathbf{r}_{ij}} u \left(\mathbf{r}_{ij} \right) \right> \ . \ee
Sometimes it is called Irving-Kirkwood formula \cite{Irving_Kirkwood_JCP_1950, BarratHansen_Liquids}.  As in the first derivation, the important point is
that this equation follows from pure mechanics, does not make any assumption related to equilibrium statistical mechanics.  In terms of
pair distributions $p^{ij}_2$ the Irving-Kirkwood formula reads
\be \begin{split} p = p_{\mathrm{ideal}} & -  \frac{N_{{\cal A}}^2}{6} \int
\mathbf{r}_{12} \cdot \partial_{{\mathbf{r}}_{12}} u^{{\cal AA}}(\mathbf{r}_{12})
p_2^{{\cal AA}}(\mathbf{r}_1, \mathbf{r}_2) -  \\ & -  \frac{N_{{\cal A}}N_{{\cal
B}}}{3} \int  \mathbf{r}_{12} \cdot \partial_{{\mathbf{r}}_{12}} u^{{\cal AB}}
(\mathbf{r}_{12})  p_2^{{\cal AB}}(\mathbf{r}_1, \mathbf{r}_2) - \\ & -
\frac{N_{{\cal B}}^2}{6} \int  \mathbf{r}_{12} \cdot \partial_{{\mathbf{r}}_{12}}
u^{{\cal BB}}(\mathbf{r}_{12})  p_2^{{\cal BB}}(\mathbf{r}_1, \mathbf{r}_2) \ .
\end{split} \ee
Using the ansatz \ref{eq:ansatz_for_p2} for $p_2$, and integrating by parts (and
remembering that $\mathbf{\nabla} \cdot \mathbf{r} = 3$), we obtain the same
result as before for the osmotic pressure.

\section{Power transfer}

In the main text, we outlined the derivation of the power transfer in the cases of
either two particles, or many particles with only pairwise collisions.  Here we
establish a more general result which suggests that the power transfer is expressed in
terms of only pair and triple correlation functions (but not higher order ones).
Consider the work performed by all forces per unit time on all ${{\cal A}}$ particles,
which is also the power received by ${{\cal A}}$ particles:
\be W=\sum_{i}^{N_{{\cal A}}}\int \frac{\partial U }{\partial \mathbf{r}_i^{{\cal
A}}} \left[\frac{T_{{\cal A}}}{\zeta_{{\cal A}}} \frac{\partial P}{\partial
\mathbf{r}_i^{{\cal A}}} + \frac{\partial U }{\partial \mathbf{r}_i^{{\cal A}}}
\frac{P}{\zeta_{{\cal A}}} \right] d \{\mathbf{r}\} \ ,
\label{eq:first_expression_for_power_to_one_A} \ee
with $U = U\left( \left\{\mathbf{r}_j^{{\cal A}} \right\} , \left\{
\mathbf{r}_k^{{\cal B}} \right\} \right)$ the total potential energy of
the system.
In the integral (\ref{eq:first_expression_for_power_to_one_A} ), the first factor
is the force which acts on particle ${{\cal A}}_i$
due to all other particles, while
the second factor (in square brackets) is the current, i.e.,
the velocity of the particle
${{\cal A}}_i$ multiplied by the probability density $P$.  Thus, the integral
(\ref{eq:first_expression_for_power_to_one_A}) is the average power transfer to
one particle ${{\cal A}}_i$.  By symmetry, it is independent of $i$, so that
the summation
over $i$ reduces to a factor $N_{{\cal A}}$.   As long as the interaction
potentials are pairwise additive (\ref{eq:pairwise_additive_interactions}), the
force is also a sum:
\be \frac{\partial U }{\partial \mathbf{r}_i^{{\cal A}}}  = \sum_{j \neq i}^{N_{{\cal A}}}\frac{\partial u^{{\cal AA}} \left( \mathbf{r}_i^{{\cal A}} - \mathbf{r}_j^{{\cal A}} \right)}{\partial \mathbf{r}_i^{{\cal A}}} + \sum_{k}^{N_{{\cal B}}} \frac{\partial u^{{\cal AB}} \left( \mathbf{r}_i^{{\cal A}} - \mathbf{r}_k^{{\cal B}} \right)}{\partial \mathbf{r}_i^{{\cal A}}} \ . \ee
The first term is the force acting on particle ${{\cal A}}_i$ due to other ${{\cal A}}$ particles, by symmetry this term vanishes on average in the sum over ${{\cal A}}$ particles.  And the second term, which is the due to ${{\cal B}}$ particles on ${{\cal A}}$ particles, yields:
\begin{widetext}
\be \begin{split}  & W = \frac{N_{{\cal A}} T_{{\cal A}}}{\zeta_{{\cal A}}} \int
\frac{\partial u^{{\cal AB}} \left( \mathbf{r} - \mathbf{r}^{\prime} \right)}
{\partial \mathbf{r}} \times \frac{\partial p_2^{{\cal AB}}\left( \mathbf{r},
\mathbf{r}^{\prime} \right) }{\partial \mathbf{r}} d^3 \mathbf{r} d^3
\mathbf{r}^{\prime}  + \frac{N_{{\cal A}}}{\zeta_{{\cal A}}} \int \left(
\frac{\partial u^{{\cal AB}} \left( \mathbf{r} - \mathbf{r}^{\prime} \right)}
{\partial \mathbf{r}}\right)^2 p_2^{{\cal AB}}\left( \mathbf{r},
\mathbf{r}^{\prime} \right)  d^3 \mathbf{r} d^3 \mathbf{r}^{\prime} + \\ & \ \ \ \
\ \ \ \ \ \  +  \frac{N^2_{{\cal A}}}{\zeta_{{\cal A}}} \int \frac{\partial
u^{{\cal AB}} \left( \mathbf{r} - \mathbf{r}^{\prime} \right)}{\partial
\mathbf{r}} \times \frac{\partial u^{{\cal AA}}\left( \mathbf{r} -
\mathbf{r}^{\prime\prime} \right) }{\partial \mathbf{r}} p_3^{{ABA}}
\left(\mathbf{r},\mathbf{r}^{\prime},\mathbf{r}^{\prime\prime} \right) d^3
\mathbf{r} d^3 \mathbf{r}^{\prime} d^3\mathbf{r}^{\prime\prime} + \\ &  \ \ \ \ \
\ \ \ \ \  + \frac{N_{{\cal A}} N_{{\cal B}}}{\zeta_{{\cal A}}} \int
\frac{\partial u^{{\cal AB}} \left( \mathbf{r}- \mathbf{r}^{\prime} \right)}
{\partial \mathbf{r}} \times \frac{\partial u^{{\cal AB}}\left( \mathbf{r} -
\mathbf{r}^{\prime\prime} \right) }{\partial \mathbf{r}} p_3^{{ABB}}
\left(\mathbf{r},\mathbf{r}^{\prime},\mathbf{r}^{\prime\prime} \right) d^3
\mathbf{r} d^3 \mathbf{r}^{\prime} d^3\mathbf{r}^{\prime\prime}
\label{eq:Power_transfer_in_terms_of_p3} \end{split} \ee

\end{widetext}
Neglecting the three body collisions (terms with $p_3$), and using the known expression (\ref{eq:ansatz_for_p2}) for $p_2$, we return to the result given in the main text.

Here, we emphasize once again that, as long as interaction potentials are pairwise
additive, as in Eq.(\ref{eq:pairwise_additive_interactions}), Eq.(\ref{eq:Power_transfer_in_terms_of_p3}) is exact.


\end{document}